\documentclass[prb,twocolumn,showpacs,floatfix,superscriptaddress]{revtex4-2}

\usepackage[utf8]{inputenc}
\usepackage[T1]{fontenc}
\usepackage[margin=2cm]{geometry}
\usepackage{float}
\usepackage{color}
\usepackage{graphicx}
\usepackage{amsmath}
\usepackage{amssymb}
\usepackage{bbm}
\usepackage{indentfirst}
\usepackage{dcolumn}
\usepackage{soul}
\usepackage{mathtools}
\usepackage{hyperref}

\begin{document}

\title{Energy-resolved tip-orbital fingerprint in scanning tunneling spectroscopy based on the revised Chen's derivative rule}

\author{Ivan Abilio}
 \email{abilio.ivan@wigner.hun-ren.hu}
 \affiliation{Institute for Solid State Physics and Optics, HUN-REN Wigner Research Center for Physics, H-1121 Budapest, Hungary}
 \affiliation{Department of Theoretical Physics, Institute of Physics, Budapest University of Technology and Economics, H-1111 Budapest, Hungary}
\author{Kriszti\'an Palot\'as}
 \email{palotas.krisztian@wigner.hun-ren.hu}
 \affiliation{Institute for Solid State Physics and Optics, HUN-REN Wigner Research Center for Physics, H-1121 Budapest, Hungary}
 \affiliation{HUN-REN-SZTE Reaction Kinetics and Surface Chemistry Research Group, University of Szeged, H-6720 Szeged, Hungary}

\date{\today}

\begin{abstract}
The revised Chen's derivative rule for electron tunneling is implemented to enable computationally efficient first-principles-based calculations of the differential conductance $dI/dV$ for scanning tunneling spectroscopy (STS) simulations. The probing tip is included through a single tip apex atom, and its electronic structure can be modeled as a linear combination of electron orbitals of various symmetries, or can be directly transferred from first-principles electronic structure calculations. By taking pristine and boron- or nitrogen-doped graphene sheets as sample surfaces, the reliability of our implementation is demonstrated by comparing its results to those obtained by the Tersoff-Hamann and Bardeen's electron tunneling models. It is highlighted that the energy-resolved direct and interference contributions to $dI/dV$ arising from the tip's electron orbitals result in a fingerprint of the particular combined surface-tip system. The significant difference between the electron acceptor boron and donor nitrogen dopants in graphene is reflected in their $dI/dV$ fingerprints. The presented theoretical method allows for an unprecedented physical understanding of the electron tunneling process in terms of tip-orbital-resolved energy-dependent $dI/dV$ maps, that is anticipated to be extremely useful for investigating the local electronic properties of novel material surfaces in the future.
\end{abstract}

\maketitle

\section{Introduction}
\label{sec:intro}

Scanning tunneling microscopy (STM) \cite{Binnig82} and its energy-resolved spectroscopic mode, scanning tunneling spectroscopy (STS) are very successful scanning probe methods to investigate nanostructured material surfaces down to (sub-)atomic scale details. Well controlled theoretical modeling through computer simulations is inevitable to understand experimental measurements of STM/STS at the  microscopic level concerning all relevant details, where achieving a good balance between accuracy and computational time efficiency is a challenge \cite{Palotas14}.

Most of the employed theoretical methods of electron tunneling in STM junctions are based on the actually calculated electronic structure of the investigated material surface, and, still, to a smaller extent of that of the probing STM tip. The mostly employed Tersoff-Hamann (TH) method considers an $s$-type tip orbital without electronic structure, i.e., with a constant electron density of states (DOS), and the tunneling current ($I$) and differential conductance ($dI/dV$) are respectively approximated as the energy-integrated and energy-resolved local DOS (LDOS) of the surface in the vacuum at the tip apex position \cite{Tersoff83,Tersoff85}. Chen's derivative rules improved the TH approximation, and electron orbitals of the STM tip beyond $s$-type were shown to correspond to proper spatial derivatives of the single electron wave functions of the investigated surface \cite{Chen90}, still without any electronic structure of the probing STM tip. The so-called revised Chen's derivative rule was introduced by M\'andi and Palot\'as, where either energy-independent or energy-dependent linear combinations of tip electron orbitals from first principles calculations were proposed, and the role of tip-orbital interference in the STM imaging was demonstrated \cite{Mandi15}. Both TH and Chen's methods can be derived from Bardeen's theory of electron tunneling \cite{Bardeen61}.

The above-listed theoretical methods assume a weak coupling between the sample surface and the STM tip, their electronic structures are usually calculated in the separate subsystems, and their effect on modifying each other's atomic geometries is neglected. Going beyond that, non-equilibrium Green function (NEGF) methods for electron transport were developed, where tunneling and contact regimes and transitions between them are all accessible within the same theoretical framework \cite{Blanco04,Blanco06,González16}. Note that NEGF methods for STM simulations are very demanding computationally since the coupled sample-tip system needs to be calculated.

Concentrating on theoretical works on STS and its simulation, the early paper of Lang demonstrated that the DOS features of both the sample surface and the tip appear in the calculated $(dI/dV)/(I/V)$ spectra \cite{Lang86}. The importance of the tip electronic structure on the $dI/dV$ tunneling spectra was highlighted in numerous publications, and some related literature was concerned with the extraction of local electronic properties of surfaces from experimental $dI/dV$ data \cite{PhysRevB.53.11176,PhysRevB.75.035421,Passoni09,PhysRevB.80.125402,PhysRevB.80.165419}. Hofer and Garcia-Lekue demonstrated the separation of tip and surface contributions to the $dI/dV$ \cite{Hofer05}. The effect of the tip magnetization and different terms contributing to the $dI/dV$ were analyzed for magnetic surfaces in spin-polarized STM and STS \cite{Palotas11,Palotas11sts,Palotas12,Palotas13}. Interorbital interference effects in STM and STS were studied by Jurczyszyn and Stankiewicz \cite{Jurczyszyn03, Jurczyszyn05}. Fingerprints of Dirac points were identified in calculated LDOS spectra of a graphene layer on metal substrates \cite{Slawinska11}. Recent major interest on STM/STS simulations is focusing on the effect of functionalized tips \cite{Krejci17,Oinonen24,Abilio24,setescak24,Paschke25,Robles25} and on twisted 2D material Moir\'e heterostructures employing simple geometry-based models \cite{LeSter24,gutierrez24}, where the inclusion of interlayer interactions and electronic structure effects is a challenge.

In the present paper we implement the revised Chen's derivative rule \cite{Mandi15} into the BSKAN code \cite{Hofer00,Hofer03,Palotas05} for the computationally efficient calculation of the tunneling differential conductance $dI/dV$, and after a careful comparison with established methods based on the TH approximation and Bardeen's theory through numerical results, we analyze the energy-resolved $dI/dV$ tip-orbital composition fingerprints of weakly coupled sample surface-tip systems by taking pristine and boron- and nitrogen-doped graphene surfaces \cite{González16,Zhao13,Zheng10,Telychko14,Telychko15,Tison15,Ferrighi15,Li18,Joucken19,Neilson19,Yang23} in combination with selected metal and functionalized tip models. As a straightforward future application of our implemented $dI/dV$ calculation method, we propose that a systematic generation of energy-resolved $dI/dV$ fingerprint datasets calculated with different methods and approximations for the tip geometric \cite{Mandi13,Mandi14,Mandi15progsurfsci} and electronic structures, in analogy to STM datasets using so-far limited parameters and data dimensions \cite{Choudhary21}, will be an asset for future data mining applications using machine learning methods for doped graphene \cite{Guerrero-Rivera24} and beyond, for other novel materials as well.

The paper is organized as follows: The revised Chen's derivative rule method applied to STM and STS simulations is given in Section \ref{sec:chen}, which is followed by describing the computational details in Section \ref{sec:comp}. The comparison of calculated $dI/dV$ data obtained by the revised Chen's method and the Tersoff-Hamann approximation and Bardeen's tunneling theory is reported and discussed in Sections \ref{sec:th} and \ref{sec:bar}, respectively. A detailed analysis of the energy-resolved tip-orbital fingerprints derived from single-point $dI/dV$ spectra is given in Section \ref{sec:orb_anal}. Finally, our conclusions are found in Section \ref{sec:conc}.

\section{Method}
\label{sec:meth}

\subsection{Revised Chen's derivative rule for STM and STS}
\label{sec:chen}

According to Bardeen's theory \cite{Bardeen61}, the tunneling current $I$ between a sample surface ($S$) and a tip ($T$) at a bias voltage $V$ applied to the tip can be calculated in the weak coupling and elastic tunneling limits as
\begin{equation}
\begin{aligned}
I(V)&=\frac{2\pi e}{\hbar}\sum_{\mu\nu}f(E_{\nu}-eV)[1-f(E_{\mu})]\\
&\times|M_{\mu\nu}|^2\delta(E_{\nu}-E_{\mu}-eV),
\end{aligned}
\label{eq:current1}
\end{equation}
where $e$ is the elementary charge, $\hbar$ is the reduced Planck constant, $f(E)$ is the Fermi-Dirac distribution function, and $|M_{\mu\nu}|^2$ is the tunneling transmission connecting two single-electron states of the sample ($S:\mu$) and the tip ($T:\nu$) through the vacuum barrier. $\mu$ and $\nu$ denote composite indices of the band ($n$), wave vector ($\mathbf{k}_{\parallel}$) and spin ($\sigma$), such that $\mu=(n^S,\mathbf{k}_{\parallel}^S,\sigma^S)$ and $\nu=(n^T,\mathbf{k}_{\parallel}^T,\sigma^T)$. The corresponding $E_{\mu}$ and $E_{\nu}$ energies are Kohn-Sham eigenenergies in the separately calculated sample ($S:\mu$) and tip ($T:\nu$) subsystems, respectively, that can be obtained by density functional theory (DFT). Here, a weak coupling is assumed, i.e., atomic relaxations in the coupled sample-tip system are neglected. The sum in Eq.~(\ref{eq:current1}) is performed over all $\mu$ sample and $\nu$ tip electron states within an energy window that corresponds to the applied bias voltage (and temperature). The tunneling transitions are assumed to be elastic, i.e., conserving energy, and this is ensured by the Dirac $\delta$ function in Eq.~(\ref{eq:current1}). Considering a finite temperature $\tilde{T}$ for the $f(E)$ functions, and approximating the Dirac $\delta$ function by a normalized Gaussian, the signed tunneling current reads
\begin{equation}
\begin{aligned}
&I(V,\tilde{T})=\frac{2\pi e}{\hbar}\mathrm{sgn}(V)\sum_{E_{\mu}\in\mathcal{W}(V,\tilde{T})}\sum_{\nu}\\&\frac{|M_{\mu\nu}|^2}{\sqrt{2\pi}k_B\tilde{T}}\exp\left[-\frac{(E_{\nu}-E_{\mu}-eV)^2}{2k_B^2\tilde{T}^2}\right],
\end{aligned}
\label{eq:current}
\end{equation}
where $\mathcal{W}(V,\tilde{T})=[\mathrm{min}(E_F^S,E_F^S+eV)-\ln(3+\sqrt{8})k_B\tilde{T},\mathrm{max}(E_F^S,E_F^S+eV)+\ln(3+\sqrt{8})k_B\tilde{T}]$ is the energy window of tunneling between sample and tip electronic states \cite{Palotas11} with $E_F^S$ being the Fermi energy of the sample surface and $k_B$ the Boltzmann constant. The sign of $I$ is determined by the sign of the bias voltage $V$ with positive and negative values corresponding to $T\rightarrow S$ (from the tip's occupied to the sample's unoccupied states) and $S\rightarrow T$ (from the sample's occupied to the tip's unoccupied states) current flow, respectively. Similarly, the differential conductance is obtained as
\begin{equation}
\begin{aligned}
&\frac{dI}{dV}(V,\tilde{T})=\lim_{\Delta V\rightarrow 0}\frac{2\pi e}{\hbar}\frac{1}{\Delta V}\sum_{E_{\mu}\in\mathcal{W}(V,\Delta V)}\sum_{\nu}\\&\frac{|M_{\mu\nu}|^2}{\sqrt{2\pi}k_B\tilde{T}}\exp\left[-\frac{(E_{\nu}-E_{\mu}-eV)^2}{2k_B^2\tilde{T}^2}\right],
\end{aligned}
\label{eq:dIdV}
\end{equation}
where the energy window now is restricted to the close vicinity of $E_F^S+eV$, such that $\mathcal{W}(V,\Delta V)=[E_F^S+e(V-\Delta V/2),E_F^S+e(V+\Delta V/2)]$. Mathematically rigorously the $\Delta V\rightarrow 0$ limit has to be taken, however, in practice the choice of a small value of $\Delta V$ is a good approximation. Note that the above formula for the direct calculation of $dI/dV$ avoids possible numerical problems of performing voltage-derivatives of the tunneling current in Eq.~(\ref{eq:current}), as also pointed out in Refs.~\cite{Hofer05,Palotas12}.

The tunneling transmission between states $\mu$ and $\nu$ is the absolute-value-square of the tunneling matrix element $M_{\mu\nu}$, which is expressed as an integral over a separation surface $\mathcal{S}$ in the vacuum as
\begin{equation}
\begin{aligned}
M_{\mu\nu}=-\frac{\hbar^2}{2m}\int_{\mathcal{S}}(\chi_{\nu}^*\boldsymbol{\nabla}\psi_{\mu}-\psi_{\mu}\boldsymbol{\nabla}\chi_{\nu}^*)\mathbf{dS},
\end{aligned}
\label{eq:m-bardeen}
\end{equation}
with $\psi_{\mu}$ and $\chi_{\nu}$ sample and tip wave functions, respectively. Evaluating the tunneling matrix elements is a computationally demanding task, especially for complex surfaces with a large number of electronic states, and modeling realistic tips is usually difficult \cite{Mandi15progsurfsci}. Chen's derivative rule provides a computationally efficient approximation for the calculation of the tunneling matrix elements \cite{Chen90}. The main idea of Chen's paper is to expand the tip wave functions $\chi_{\nu}$ into real spherical harmonic components $\beta\in\{s,p_y,p_z,p_x,d_{xy},d_{yz},d_{3z^2-r^2},d_{xz},d_{x^2-y^2}\}$ centered at the tip apex atomic site position $\mathbf{r}_0$. According to this, and using the notation of M\'andi and Palot\'as \cite{Mandi15}, the individual contributions of specific $\beta$ tip electron orbitals are obtained as spatial derivatives of the sample wave functions at $\mathbf{r}_0$,
\begin{equation}
\begin{aligned}
M_{\mu\nu\beta}=\frac{2\pi\hbar^2}{\kappa_\nu m}C_{\nu\beta}\,\hat{\partial}_{\nu\beta}\,\psi_{\mu}(\mathbf{r}_0).
\end{aligned}
\label{eq:m_munubeta}
\end{equation}
Here, $\kappa_{\nu}$ is the vacuum decay of the tip wave function $\chi_{\nu}$, $m$ is the electron's mass, $C_{\nu\beta}$ are proper coefficients, which are, in general, energy-dependent complex or real numbers (see Sec.~II.B of Ref.~\cite{Mandi15}), and the differential operators $\hat{\partial}_{\nu\beta}$ for the $\beta$ orbital characters \cite{Chen90} are listed in Table \ref{tab_chen}. A detailed description of obtaining the spatial derivatives of $\psi_{\mu}(\mathbf{r})$ assuming a plane-wave expansion is reported in Sec.~II.A of Ref.~\cite{Mandi15}. Note that, in general, arbitrary tip orientations \cite{Mandi13,Mandi14,Mandi15progsurfsci} can also be included into the revised Chen's electron tunneling model by redefining the $C_{\nu\beta}$ coefficients upon rigidly rotating the electron orbital structure of the tip apex atom, for the detailed formalism see Sec.~II.C of Ref.~\cite{Mandi15}, and applications of this approach are reported for STM in Refs.~\cite{Mandi15,Abilio24}.

\begin{table*}
\caption{Differential operators $\hat{\partial}_{\nu\beta}$ for real-space electron orbital characters $\beta$ according to Chen \cite{Chen90}, where $\kappa_{\nu}$ is the vacuum decay of the tip wave function $\chi_{\nu}$.}
\begin{ruledtabular}
\begin{tabular}{cccccccccc}
$\beta$ & $s$ & $p_y$ & $p_z$ &  $p_x$ & $d_{xy}$ & $d_{yz}$ & $d_{3z^2-r^2}$ & $d_{xz}$ & $d_{x^2-y^2}$\\
\hline
$\hat{\partial}_{\nu\beta}$ & 1 & $\kappa_{\nu}^{-1}\frac{\partial}{\partial y}$ & $\kappa_{\nu}^{-1}\frac{\partial}{\partial z}$ & $\kappa_{\nu}^{-1}\frac{\partial}{\partial x}$ & $\kappa_{\nu}^{-2}\frac{\partial^2}{\partial x\partial y}$ & $\kappa_{\nu}^{-2}\frac{\partial^2}{\partial y\partial z}$ & $3\kappa_{\nu}^{-2}\frac{\partial^2}{\partial z^2}-1$ & $\kappa_{\nu}^{-2}\frac{\partial^2}{\partial x\partial z}$ & $\kappa_{\nu}^{-2}\left(\frac{\partial^2}{\partial x^2}-\frac{\partial^2}{\partial y^2}\right)$
\end{tabular}
\end{ruledtabular}
\label{tab_chen}
\end{table*}

Using the above notation, the tunneling matrix elements can be written as the sum of the real spherical harmonic components, $M_{\mu\nu}=\sum_{\beta}M_{\mu\nu\beta}$, and the tunneling transmission can be expressed in terms of the tip orbitals $\beta$ as \cite{Mandi15}
\begin{equation}
\begin{aligned}
|M_{\mu\nu}|^2&=\left|\sum_\beta M_{\mu\nu\beta}\right|^2=\sum_{\beta,\beta'}M_{\mu\nu\beta}^*M_{\mu\nu\beta'}\\
&=\sum_\beta|M_{\mu\nu\beta}|^2+\sum_{\beta\not=\beta'}2\mathrm{Re}(M_{\mu\nu\beta}^*M_{\mu\nu\beta'}).
\end{aligned}
\label{eq:m2-decomp}
\end{equation}
This partitioning of the transmission enables the expressions of the current in Eq.~(\ref{eq:current}) and of the differential conductance in Eq.~(\ref{eq:dIdV}) in terms of tip-orbital contributions,
\begin{equation}
\begin{aligned}
I&=\sum_{\beta\beta'}I_{\beta\beta'}=\sum_{\beta}I_{\beta\beta}+\sum_{\beta\ne\beta'}I_{\beta\beta'},\\
\frac{dI}{dV}&=\sum_{\beta\beta'}\left(\frac{dI}{dV}\right)_{\beta\beta'}=\sum_{\beta}\left(\frac{dI}{dV}\right)_{\beta\beta}+\sum_{\beta\ne\beta'}\left(\frac{dI}{dV}\right)_{\beta\beta'},
\end{aligned}
\label{eq:decomp}
\end{equation}
where the notations of the bias-voltage- and temperature-dependence are omitted for brevity. $I_{\beta\beta'}=I_{\beta'\beta}$ and $(dI/dV)_{\beta\beta'}=(dI/dV)_{\beta'\beta}$ represent real-valued symmetric current composition \cite{Abilio24} and differential conductance composition matrices, respectively. In general, these matrices have the dimensions of $(2l+1)\times(2l+1)$ with $l$ being the angular quantum number, and as Table \ref{tab_chen} indicates, our implementation is restricted to $l_\mathrm{max}=2$, resulting in matrices with a maximal dimension of $9\times 9$. The diagonal elements of these matrices correspond to positive contributions to $I$ or $dI/dV$ in terms of tip orbitals, and the off-diagonal elements to tip-orbital interference contributions with positive or negative values meaning constructive or destructive interference, respectively. The analysis of the $I_{\beta\beta'}$ and $(dI/dV)_{\beta\beta'}$ matrix elements and their percentual contribution to the total $I$ and total $dI/dV$ provides a deeper physical understanding of the electron tunneling process in terms of the contributing tip electron orbital characters. Such an analysis for the current $I$ has been provided earlier \cite{Mandi15,Abilio24}, and the present paper proposes and demonstrates that it is also possible for the differential conductance $dI/dV(V)$ in high voltage/energy-resolution. In the latter case the percentual $dI/dV$ composition matrices are
\begin{equation}
\begin{aligned}
\left(\frac{d\tilde{I}}{dV}\right)_{\beta\beta'}=\left(\frac{dI}{dV}\right)_{\beta\beta'}/\sum_{\beta\beta'}\left(\frac{dI}{dV}\right)_{\beta\beta'}
\end{aligned}
\label{eq:dIdV-relative}
\end{equation}
with the sum of all such matrix elements $\sum_{\beta\beta'}(d\tilde{I}/dV)_{\beta\beta'}$ being 100\%. Moreover, note that destructive tip-orbital interference terms appear as negative numbers, thus in the presence of such terms the sum of all other contributions exceeds 100\% using the above definition.

\subsection{Computational details}
\label{sec:comp}

The above formalism has been used to calculate $dI/dV(V)$ spectra and $dI/dV(x,y,V)$ maps of pure and boron- and nitrogen-doped graphene single layers, which will be denoted by GR, GB, and GN, respectively, in the following. The electronic states of these systems were calculated by non-spin-polarized DFT using the Vienna Ab-initio Simulation Package \cite{vasp}. In practice, $dI/dV$ in Eq.~(\ref{eq:dIdV}) is evaluated at a small $\Delta V$ value, in our case, $\Delta V$=10 meV, which also equals our voltage resolution in the bias voltage range of [-1 V, +1 V], resulting in 201 voltage points. Since $\Delta V$ is small enough, in the following the tunneling probability expressed by the normalized Gaussian function in Eq.~(\ref{eq:dIdV}) is replaced by unity.

The pure (GR) and doped (GB, GN) graphene systems are modeled based on a free-standing single layer graphene sheet in a $7\times 7$ surface unit cell with 98 carbon atoms \cite{Telychko14,Mandi15} and a 24-\AA-wide vacuum region perpendicular to the surface to avoid unphysical interactions between the repeated neighboring graphene layers. For the substitutionally doped graphene sheets one carbon atom is replaced either by a boron (for GB) or a nitrogen atom (for GN) in the given supercell, resulting in about 1\% doping concentration. The exchange-correlation functional parametrized by Perdew and Wang \cite{PW91} within the generalized gradient approximation and the projector augmented wave method \cite{paw} were used with an energy cutoff of 400 eV for the plane-wave expansion of the electron wave functions, and an $11\times 11\times 1$ Monkhorst-Pack \cite{Monkhorst76} k-point sampling of the Brillouin zone. The convergence criterion of 0.01 eV\AA$^{-1}$\, for atomic forces was used in the structural optimizations, which resulted in planar atomic sheet structures in all cases.

\begin{figure}[ht!]
\includegraphics[width=\columnwidth]{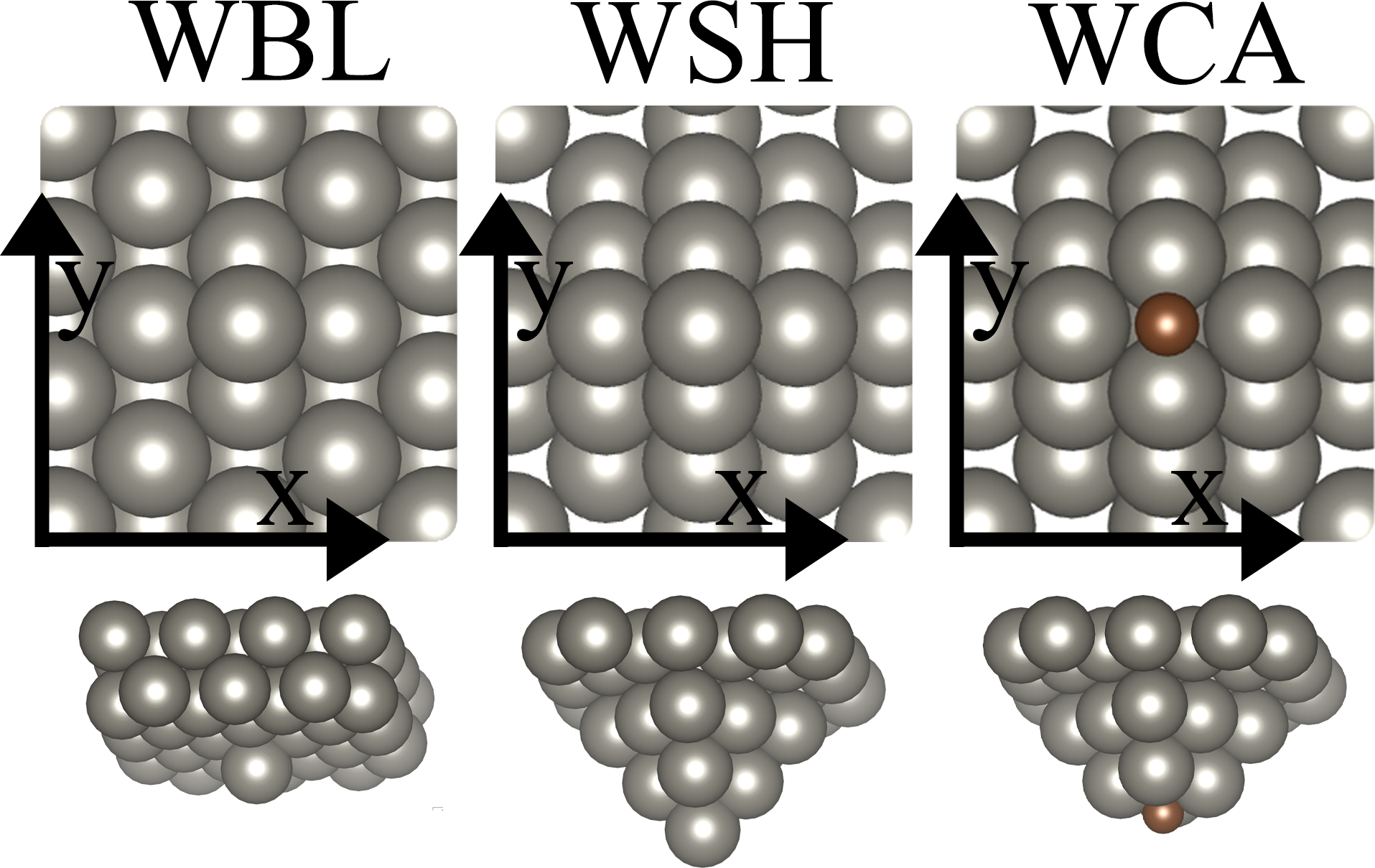}
\caption{Atomic geometries of the used tungsten-based tip models: $\mathrm{W_{blunt}}$ (WBL), $\mathrm{W_{sharp}}$ (WSH), and $\mathrm{W_{C-apex}}$ (WCA). Top and perspective views are presented. Tungsten and carbon atoms are represented by gray and brown spheres, respectively.}
\label{fig0}
\end{figure}

Several model tips are considered for the revised Chen’s derivative rule: (i) idealized model tips of pure $s$, pure $p_z$, and a combination of $(s+p_z)/\sqrt{2}$ orbital characters with constant density of states, for our results see Sec.~\ref{sec:th}, and (ii) three tungsten tip models with different apex structures and compositions: $\mathrm{W_{blunt}}$, $\mathrm{W_{sharp}}$, and $\mathrm{W_{C-apex}}$, which will be denoted by WBL, WSH and WCA, respectively, see Secs.~\ref{sec:bar} and \ref{sec:orb_anal}. As shown in Figure \ref{fig0} the WBL tip model consists of a tungsten adatom adsorbed on the hollow site of a W(110) surface slab, the WSH tip has a tungsten atomic pyramid of three-atoms height on a W(110) surface slab, and the WCA tip is obtained from the WSH tip by replacing its tungsten apex atom by a carbon atom (and the atomic structure reoptimized) to model a deliberately carbon-functionalized tip or a possible carbon contamination from the sample graphene sheet. More details on the used tip geometries and electronic structures are reported elsewhere \cite{Teobaldi12}.

\section{Results and Discussion}
\label{sec:res}

To demonstrate the applicability of the implemented revised Chen's STS method, pure and doped graphene sheets are investigated. First, the obtained results are compared with those of already established electron tunneling models, i.e., the Tersoff-Hamann approximation and Bardeen's model, and then the energy-resolved tip-orbital composition of the differential conductance is analyzed.

\subsection{Comparison with the Tersoff-Hamann method}
\label{sec:th}

\begin{figure}[ht!]
\includegraphics[width=\columnwidth]{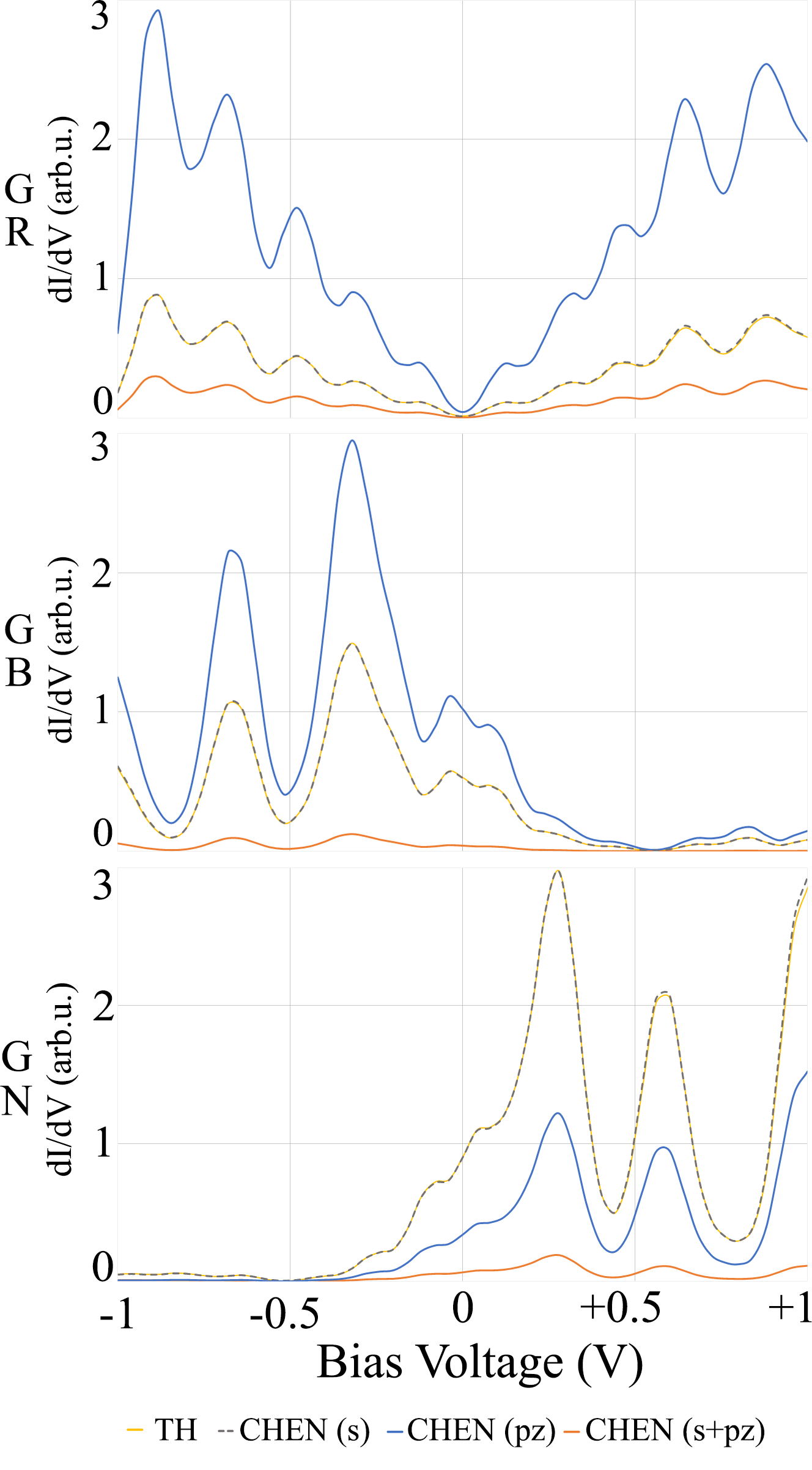}
\caption{Differential conductance $dI/dV(V)$ spectra above a carbon atom in pure graphene (GR, top), and above the substitutional (graphitic) defect atom in boron-doped (GB, middle) and nitrogen-doped (GN, bottom) graphene. Calculated $dI/dV(V)$ spectra obtained by the Tersoff-Hamann method (TH, yellow) and by revised Chen's derivative rules (CHEN) with $s$, $p_z$ and $(s+p_z)/\sqrt{2}$ tip orbitals (dashed gray, blue, and orange, respectively) are compared at a tip-sample distance of 7 \AA. The $dI/dV$ values obtained by CHEN are rescaled so that the highest peak values obtained by CHEN(s) and TH are matching exactly. Note that the $dI/dV$ scales of the three subfigures are not comparable.}
\label{fig1}
\end{figure}

Figure \ref{fig1} shows calculated $dI/dV(V)$ spectra for the GR, GB, and GN systems at a constant-height condition obtained by the Tersoff-Hamann (TH) model and the revised Chen's method (CHEN) for different tip orbitals: pure $s$, pure $p_z$, and their combination $(s+p_z)/\sqrt{2}$. The main purpose is to compare the spectra in relation to the results of the simple Tersoff-Hamann model, and to study the effect of the different tip orbitals on the $dI/dV$ values.

For GR we observe the typical V-shaped spectra centered at zero bias voltage, which corresponds to the energetic position of the Dirac cone. On the other hand, the B- and N-dopings change the shape of the $dI/dV$ curves to become asymmetric, and the bottom of the spectra is shifted along the voltage axis in opposite directions by about 0.5 V \cite{Zheng10} according to their respective electron acceptor (B) and donor (N) character with respect to the substituted carbon atom \cite{Ferrighi15,Joucken19}.

As expected, the spectra obtained by TH and CHEN(s) coincide for each case, which validates the correctness of our implementation. The scaling factor applied to the CHEN(s) results in relation to TH can be used to express the $dI/dV$ values obtained by CHEN for any tip orbital compositions in the units of those obtained by TH (\AA$^{-3}V^{-1}$ units). Furthermore, it is found that a $p_z$ tip orbital increases the $dI/dV$ values in comparison to an $s$ tip orbital for GR and GB only, and GN exhibits an opposite behavior. In general, this can be understood by the enlarged wave function overlap of sample electronic states with an oriented $p_z$ tip orbital pointing toward the surface for GR and GB. However, GN behaves differently due to the presence of a strong (destructive) interference of the surface local electronic structure around the N atom \cite{Telychko14,Telychko15}, which is even more enhanced by the more oriented probing $p_z$ tip orbital. When using an $(s+p_z)/\sqrt{2}$ tip orbital composition, the $dI/dV$ values are consistently reduced in the full bias voltage range in comparison to either an $s$ or a $p_z$ tip orbital, which is due to the destructive tip-orbital interference also reported for the STM image analysis of N-doped graphene \cite{Mandi15}. Note that the $dI/dV$ peak and dip positions do not change in energy, irrespectively of the employed electronically featureless tip orbital characters, and they reflect the underlying electronic structure of the sample surface. The obtained characteristic peak at -0.35 V for GB is in agreement with an increased $dI/dV$ reported in Ref.~\cite{Telychko15}. The characteristic peaks at about +0.25 V and +0.60 V in the GN spectrum can be attributed to the corresponding peaks of the projected DOS of the N dopant atom, as reported in Ref.~\cite{Telychko14}.

\begin{figure}[ht!]
\includegraphics[width=\columnwidth]{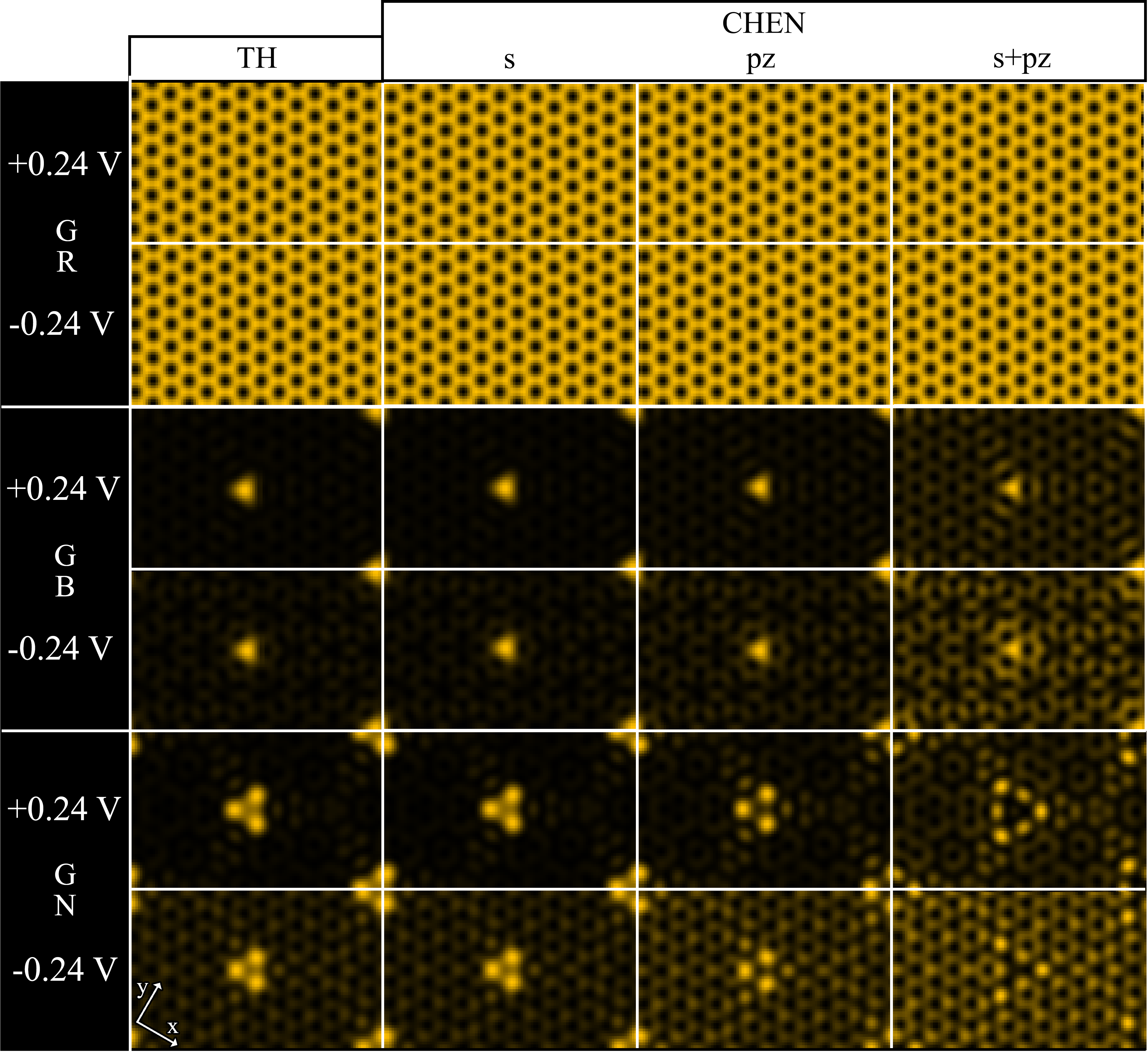}
\caption{Differential conductance $dI/dV(x,y,V)$ maps for pure (GR, top 2 rows), boron-doped (GB, middle 2 rows) and nitrogen-doped (GN, bottom 2 rows) graphene at bias voltages of $V=\pm 0.24$ V. Calculated 2D $dI/dV(x,y)$ maps obtained by the Tersoff-Hamann method (TH) and by revised Chen's derivative rules (CHEN) with $s$, $p_z$ and $(s+p_z)/\sqrt{2}$ tip orbitals (denoted by s, pz, and s+pz, respectively) are compared at a constant tip-sample distance of 7 \AA. The individual image size is 29.86 \AA $\times$ 17.24 \AA, which was calculated on a $101\times 58$ lateral grid and $\approx 0.3$ \AA\, spatial resolution. The substitutional (graphitic) defects are in the center of the images for GB and GN. The coordinate system is shown in the bottom left.}
\label{fig2}
\end{figure}

To demonstrate the STS simulation capability of our method, 2D $dI/dV(x,y)$ maps of GR, GB, and GN are shown in Figure \ref{fig2} at bias voltages of $\pm 0.24$ V at a constant-height condition. As expected, the comparison between STS maps obtained by TH and CHEN(s) methods results in identical images for the corresponding surfaces. When comparing the effect of the tip orbitals $s$, $p_z$ and the composition of $(s+p_z)/\sqrt{2}$, we do not find any significant difference for GR, and the honeycomb structure is resolved as a bright network in the STS maps at the selected voltages. The STS maps of GB consistently show bright protrusions above the B dopant atom, in agreement with STM images reported in Refs.~\cite{Zhao13,Telychko15}, and the background (carbon) features are more pronounced when imaged with an $(s+p_z)/\sqrt{2}$ tip orbital composition, and also at a negative voltage. The latter two findings are also true for GN, but the STS contrast character of the N dopant atom is different than seen for the B dopant, as also reported for STM images in Ref.~\cite{Telychko15}. The nitrogen position appears as a dark dip and the three neighboring carbon atoms are resolved as bright spots with pure $s$ and pure $p_z$ tip orbitals. However, the contrast is further changed when an $(s+p_z)/\sqrt{2}$ tip orbital composition is used, and the nitrogen's third nearest neighbor carbons appear even brighter than the nearest neighbor carbons in the STS maps, while the dip above the nitrogen is more extended in size. These findings for the STS maps of GN in Fig.~\ref{fig2} are consistent with STM experiments \cite{Telychko14,Telychko15} and with a similar analysis of STM images of the tunneling current employing the revised Chen's derivative rules \cite{Mandi15}, where a destructive interference between the $s$ and $p_z$ tip orbitals was identified.

\subsection{Comparison with the Bardeen method}
\label{sec:bar}

\begin{figure}[ht!]
\includegraphics[width=\columnwidth]{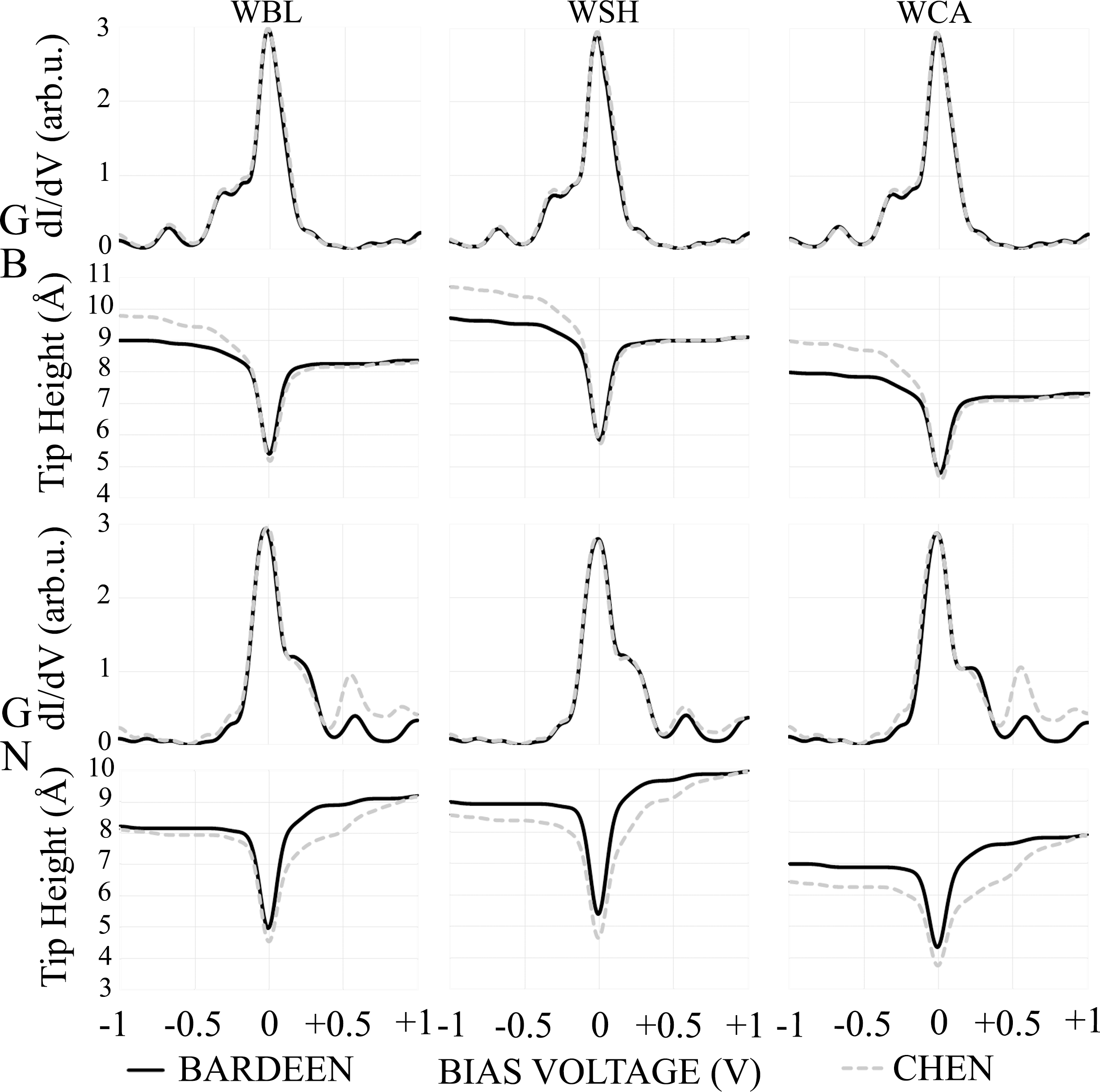}
\caption{Differential conductance $dI/dV(V)$ spectra above the substitutional (graphitic) defect atom in boron-doped (GB, first row) and nitrogen-doped (GN, third row) graphene. Calculated $dI/dV(V)$ spectra obtained by the Bardeen method (BARDEEN, black) and by the revised Chen's method (CHEN, dashed gray) using a blunt tungsten tip (WBL), a sharp tungsten tip (WSH) and a sharp tungsten tip with a carbon apex atom (WCA) are compared. The $dI/dV(V)$ spectra are reported at a constant-current condition ($I=$10 nA) and the tip-sample distance (tip height) is reported accordingly (GB: second row, GN: fourth row). The $dI/dV$ values obtained by CHEN are rescaled to match the highest peak obtained by BARDEEN for the given surface-tip combination. The $dI/dV$ scale is the same for all $dI/dV$ subfigures.}
\label{fig3}
\end{figure}

Figure \ref{fig3} reports calculated $dI/dV(V)$ spectra for the GB and GN systems at a constant-current condition obtained by the Bardeen model and the revised Chen's method using three tungsten tip models: WBL, WSH, and WCA, for their details see Sec.~\ref{sec:comp}. Here, the main purpose is to compare the spectra in relation to the results of the Bardeen model, and study the change of the tip-sample distance (tip height) keeping the current constant.

Each spectrum in Fig.~\ref{fig3} shows the highest peak below and very close to zero bias voltage with a concomitant shift of the tip height to its lowest value. The oppositely asymmetric shape of the spectra for GB and GN is also visible, where the spectra for GB exhibit more prominent features at negative voltages, while for GN at positive voltages, in accordance with Fig.~\ref{fig1}. Overall, the pairwise agreement between the spectra obtained by the Bardeen and revised Chen's methods is very good, and the characteristic energy positions of the $dI/dV$ peaks and dips also agree. This is absolutely clear for GB, and there are slight deviations occurring for GN at larger positive voltages. In the latter case the $dI/dV$ values also vary considerably. Again, we attribute these findings for GN to the presence of a strong (destructive) interference of the surface local electronic structure around the N atom \cite{Telychko14,Telychko15}, and our results demonstrate that the employed tunneling model also shows sensitivity to such complex surface electronic structure effects.

Next, let us analyze the tip heights following the constant current value of $I=$10 nA applied in Fig.~\ref{fig3}. For GB it can be observed that the pairwise good agreement of the $dI/dV$ spectra obtained by the Bardeen and revised Chen's methods is found for almost equal tip heights at positive bias voltages. However, the tip height is varied depending on the actual tip model, and the following approximated values are obtained above +0.20 V: 8 \AA\, (WBL), 9 \AA\,(WSH), and 7 \AA\,(WCA). This trend clearly reflects that the sharp tungsten tip (WSH) provides the most localized current due to its apex structure, and that tunneling through the carbon apex atom in case of the WCA tip lowers the current \cite{Teobaldi12}, which is compensated by reducing the tip height to keep the current constant. At negative voltages, where the more prominent $dI/dV$ features occur for GB, it is found that the tip heights are systematically larger than obtained at positive voltages. This is consistent with Fig.~\ref{fig1}, where larger $dI/dV$ (LDOS for $s$ tip orbital) is visible at negative voltages for GB, and $dI/dV$ measurements also suggested this \cite{Telychko15}. The increased current needs to be compensated by increasing the tip height to keep the pre-set constant current value. It is also found for GB at negative voltages that the tip height is larger by about 1 \AA\, using Chen's rather than Bardeen's method. In this case the revised Chen's method overestimates the current compared to Bardeen's method at a given tip-sample distance.

In Fig.~\ref{fig3} we see an opposite behavior for the tip heights at constant current for GN than for GB, namely for GN the tip heights are larger at positive rather than at negative voltages, which is in line with the corresponding $dI/dV$ (LDOS for $s$ tip orbital) characteristics shown in Fig.~\ref{fig1}. Furthermore, the tip height is systematically lower in the full voltage range as calculated by Chen's rather than Bardeen's method, indicating that for GN the revised Chen's method underestimates the current compared to Bardeen's method at a given tip-sample distance, which is compensated by reducing the tip height to keep the current constant. Note that by adjusting the tip height in the positive voltage range, a closer match between the results of the two tunneling methods for GN could be achieved both in terms of tip heights and in the $dI/dV$ values.

\begin{figure*}[t!]
\includegraphics[width=\textwidth]{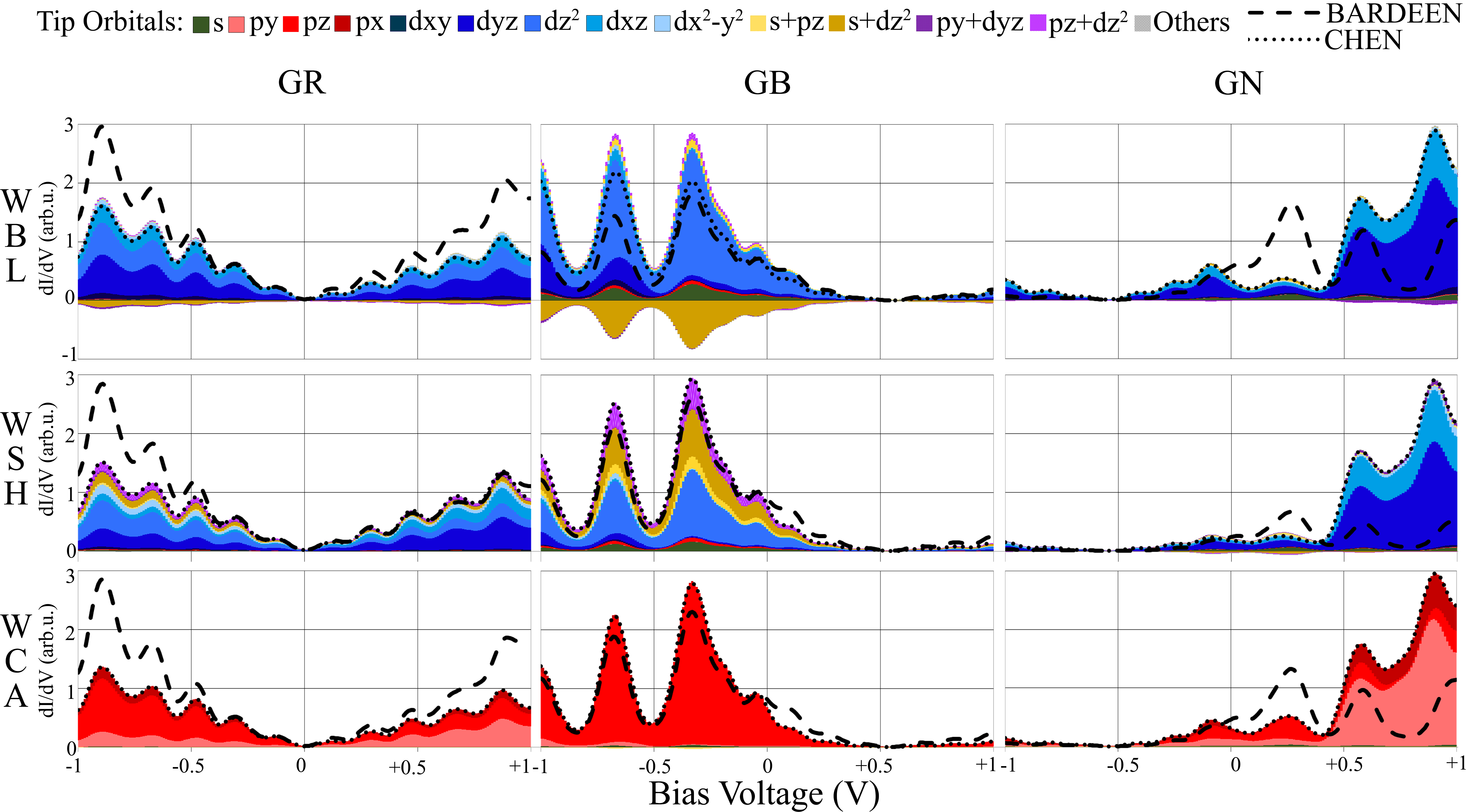}
\caption{Differential conductance $dI/dV(V)$ spectra and their tip-orbital composition above a carbon atom in pure graphene (GR, left column), and above the substitutional (graphitic) defect atom in boron-doped (GB, middle column) and nitrogen-doped (GN, right column) graphene. Calculated $dI/dV(V)$ spectra obtained by Bardeen's method (BARDEEN, black dashed) and by the revised Chen's method (CHEN, black dotted) using a blunt tungsten tip (WBL, top row), a sharp tungsten tip (WSH, middle row) and a sharp tungsten tip with a carbon apex atom (WCA, bottom row) are compared at a constant tip-sample distance of 7 \AA. The $dI/dV$ values obtained by CHEN are rescaled to match that of BARDEEN at -0.04 V. The tip-orbital composition of $dI/dV$ is given in terms of the revised Chen's derivative rules, Eq.~(\ref{eq:decomp}). The direct tip-orbital contributions and the largest tip-orbital interference contributions to $dI/dV$ are explicitly indicated as colored histograms, and the smaller interference contributions are grouped and denoted as Others. Note that the $d_{3z^2-r^2}$ tip orbital is denoted by dz$^2$ in the legend.}
\label{fig4}
\end{figure*}

Figure \ref{fig4} shows calculated $dI/dV(V)$ spectra for the GR, GB and GN systems at a constant-height condition obtained by the Bardeen model and the revised Chen's method using the three considered tungsten tip models: WBL, WSH, and WCA. All spectra obtained by the two tunneling models are matched pairwisely at the same energetic position as for Fig.~\ref{fig3}. Additionally, the energy-dependent composition of the $dI/dV$ is reported in terms of the tip orbitals and their interferences following Eq.~(\ref{eq:decomp}). The tip-orbital contributions are given by colored histograms.

We find that all calculated spectra for the separate surface structures GR, GB and GN in the respective columns of Fig.~\ref{fig4} and the corresponding spectra calculated by electronically featureless $s$, $p_z$ and $(s+p_z)/\sqrt{2}$ tip models in Fig.~\ref{fig1} are qualitatively very similar, but their quantitative details and for the tungsten tip models their tip-orbital compositions differ considerably depending on the actual tip model used. The detailed analysis of the tip orbital contributions to the $dI/dV$ spectra is reported in the next section.

We can observe in Fig.~\ref{fig4} that for GR the $dI/dV$ values are generally lower when using the revised Chen's method compared to Bardeen's. In the case of GB, Chen's method provides larger $dI/dV$ values in the negative bias voltage range compared to Bardeen's method, which results in larger current values when integrated in this range, in agreement with the finding for the tip height at constant current in Fig.~\ref{fig3}. For GN the main observation is that the peak at about +0.25 V is much more dominant when using Bardeen's method compared to Chen's, and this is true for all considered tungsten tip models. As shown in Ref.~\cite{Telychko14} the projected DOS of the nitrogen atom has its dominant peak exactly at this energetic position. This implies that Bardeen's method is more sensitive to this energetically localized state, and the different spatial derivatives of the wave function of this state following the actual tip orbital composition in the spirit of Chen's derivative rules somewhat suppress the $dI/dV$ value here.

The reason of the identified differences in the $dI/dV$ spectra lies within the construction of the used tunneling models. While the revised Chen's method uses the tip orbital electronic structure of a single tip apex atom to obtain the (current and) differential conductance values, Bardeen's method collects all contributions from overlapping single electron states between the sample and the tip. For example, for the actual tip models WBL, WSH and WCA, the contributions stemming from the four nearest neighbor atoms of the tip apex atom are zero by using Chen's derivative rules, but they are not neglected by using Bardeen's model. This fundamental difference of the calculated $dI/dV$ quantities is visible in Fig.~\ref{fig4}.

Let us now turn to the computational times of the employed electron tunneling models. After performing many tests taking the GB-WBL surface-tip combination as an example, we find that the computational advantage of the revised Chen's method is apparent when used for calculating STS maps, and the better the spatial resolution of the STS images the more efficient is using the revised Chen's method compared to Bardeen's. For a $101\times 58$ lateral grid (such as used for the subfigures in Fig.~\ref{fig2}) and 201 bias voltage points, the revised Chen's method is faster by 37 times than Bardeen's method.

\subsection{Analysis of the tip-orbital composition of the differential conductance}
\label{sec:orb_anal}

\begin{figure*}[t!]
\includegraphics[width=\textwidth]{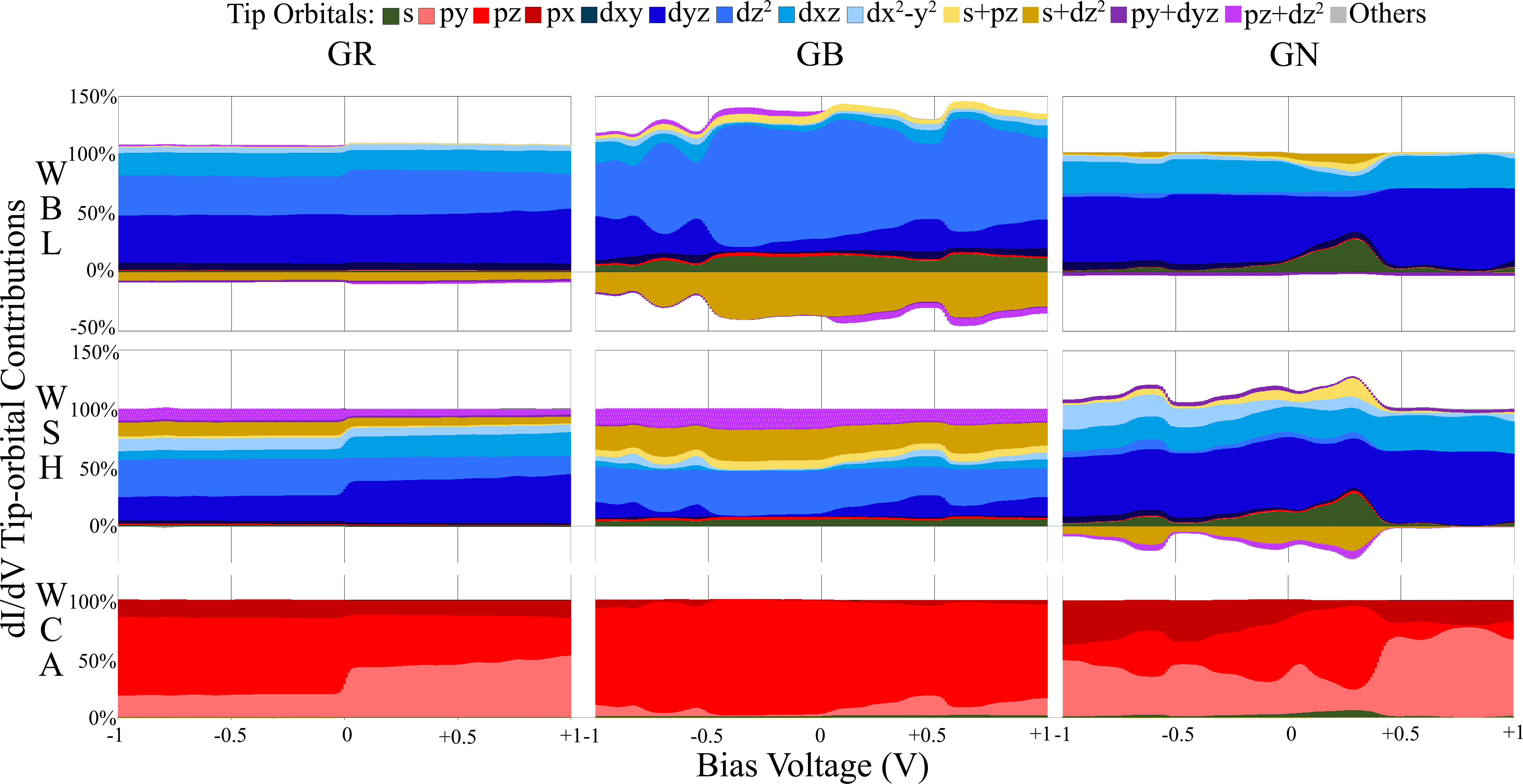}
\caption{Percentual tip orbital contributions to the differential conductance ($d\tilde{I}/dV$, Eq.~(\ref{eq:dIdV-relative})) obtained by the revised Chen's derivative rules for the studied surface-tip combinations reported in Fig.~\ref{fig4}. The direct tip-orbital contributions and the largest tip-orbital interference contributions to $dI/dV$ are explicitly indicated as colored histograms, and the smaller interference contributions are grouped and denoted as Others. Visible deviations from 100\% are due to the presence of significant negative (destructive) tip-orbital interference contributions, which have to be compensated by other tip orbitals to reach 100\%, according to the definition given in Eq.~(\ref{eq:dIdV-relative}). Note that the $d_{3z^2-r^2}$ tip orbital is denoted by dz$^2$ in the legend.}
\label{fig5}
\end{figure*}

Complementing the visualization of the absolute and relative tip orbital contributions to the $dI/dV$ respectively presented in Figs.~\ref{fig4} and \ref{fig5}, Tables \ref{tabf}, \ref{tabm} and \ref{tabp} report quantified direct tip-orbital and combined tip-orbital interference contribution values by averaging them in the full, negative, and positive bias voltage ranges, respectively. In the following discussion we refer to these Figures and Tables when making our statements.

\begin{table*}[b]
\caption{Full voltage range percentual tip-orbital contributions to the differential conductance ($d\tilde{I}/dV$, Eq.~(\ref{eq:dIdV-relative})) for the studied surface-tip combinations reported in Figs.~\ref{fig4} and \ref{fig5}. The direct tip-orbital contributions and the sum of the tip-orbital interferences are shown in the full calculated bias voltage range from -1 V to +1 V. The two largest direct tip-orbital contributions for each surface-tip combination are shown in boldface. The sum of each row equals 100\%.}
\begin{ruledtabular}
\begin{tabular}{ccccccccccc}
& $s$ & $p_y$ & $p_z$ &  $p_x$ & $d_{xy}$ & $d_{yz}$ & $d_{3z^2-r^2}$ & $d_{xz}$ & $d_{x^2-y^2}$ & Interferences\\
\hline
GR-WBL & 1.3\% & 0.1\% & 0.4\% & 0.1\% & 6.0\% & \textbf{41.7\%} & \textbf{34.4\%} & 18.9\% & 4.7\% & -7.6\% \\
GB-WBL & 11.6\% & 0.1\% & 1.9\% & 0.0\% & 4.0\% & \textbf{18.4\%} & \textbf{80.2\%} & 8.3\% & 3.1\% & -27.6\% \\
GN-WBL & 6.4\% & 0.2\% & 0.3\% & 0.1\% & 4.4\% & \textbf{55.8\%} & 2.1\% & \textbf{25.1\%} & 3.3\% & 2.3\% \\
GR-WSH & 1.0\% & 0.1\% & 0.7\% & 0.0\% & 2.2\% & \textbf{29.3\%} & \textbf{24.9\%} & 13.3\% & 8.5\% & 20.0\% \\
GB-WSH & 5.3\% & 0.0\% & 1.9\% & 0.0\% & 0.9\% & \textbf{8.6\%} & \textbf{31.7\%} & 3.9\% & 3.5\% & 44.2\% \\
GN-WSH & 7.3\% & 0.2\% & 0.7\% & 0.0\% & 2.8\% & \textbf{55.3}\% & 3.2\% & \textbf{23.5\%} & 10.6\% & -3.6\% \\
GR-WCA & 0.5\% & \textbf{32.3\%} & \textbf{53.2\%} & 14.2\% & - & - & - & - & - & -0.2\%  \\
GB-WCA & 1.7\% & \textbf{6.9\%} & \textbf{89.1\%} & 2.9\% & - & - & - & - & - & -0.6\% \\
GN-WCA & 2.2\% & \textbf{45.6\%} & \textbf{30.1\%} & 22.5\% & - & - & - & - & - & -0.4\%
\end{tabular}
\end{ruledtabular}
\label{tabf}
\end{table*}

\begin{table*}
\caption{Negative voltage range percentual tip-orbital contributions to the differential conductance ($d\tilde{I}/dV$, Eq.~(\ref{eq:dIdV-relative})) for the studied surface-tip combinations reported in Figs.~\ref{fig4} and \ref{fig5}. The direct tip-orbital contributions and the sum of the tip-orbital interferences are shown in the negative calculated bias voltage range from -1 V to 0 V. The two largest direct tip-orbital contributions for each surface-tip combination are shown in boldface. The sum of each row equals 100\%.}
\begin{ruledtabular}
\begin{tabular}{ccccccccccc}
& $s$ & $p_y$ & $p_z$ &  $p_x$ & $d_{xy}$ & $d_{yz}$ & $d_{3z^2-r^2}$ & $d_{xz}$ & $d_{x^2-y^2}$ & Interferences\\
\hline
GR-WBL & 1.2\% & 0.1\% & 0.4\% & 0.1\% & 6.0\% & \textbf{41.0\%} & \textbf{33.0\%} & 19.9\% & 4.7\% & -6.4\% \\
GB-WBL & 10.2\% & 0.0\% & 2.2\% & 0.0\% & 3.3\% & \textbf{17.8\%} & \textbf{76.8\%} & 8.6\% & 2.6\% & -21.5\% \\
GN-WBL & 3.3\% & 0.1\% & 0.2\% & 0.1\% & 5.2\% & \textbf{56.4\%} & 2.5\% & \textbf{27.0\%} & 4.0\% & 1.2\% \\
GR-WSH & 1.2\% & 0.1\% & 0.9\% & 0.0\% & 2.6\% & \textbf{21.1\%} & \textbf{31.3\%} & 7.9\% & 10.0\% & 24.9\% \\
GB-WSH & 4.9\% & 0.0\% & 2.1\% & 0.0\% & 0.8\% & \textbf{5.3\%} & \textbf{34.5\%} & 2.0\% & 3.2\% & 47.2\% \\
GN-WSH & 6.1\% & 0.2\% & 0.7\% & 0.0\% & 4.3\% & \textbf{53.7}\% & 4.7\% & \textbf{20.1\%} & 16.0\% & -5.8\% \\
GR-WCA & 0.5\% & \textbf{19.1\%} & \textbf{65.7\%} & 15.2\% & - & - & - & - & - & -0.5\%  \\
GB-WCA & 1.3\% & \textbf{3.6\%} & \textbf{93.1\%} & 2.9\% & - & - & - & - & - & -0.9\%  \\
GN-WCA & 1.9\% & \textbf{38.0\%} & \textbf{30.5\%} & 30.2\% & - & - & - & - & - & -0.6\%
\end{tabular}
\end{ruledtabular}
\label{tabm}
\end{table*}

\begin{table*}
\caption{Positive voltage range percentual tip-orbital contributions to the differential conductance ($d\tilde{I}/dV$, Eq.~(\ref{eq:dIdV-relative})) for the studied surface-tip combinations reported in Figs.~\ref{fig4} and \ref{fig5}. The direct tip-orbital contributions and the sum of the tip-orbital interferences are shown in the positive calculated bias voltage range from 0 V to +1 V. The two largest direct tip-orbital contributions for each surface-tip combination are shown in boldface. The sum of each row equals 100\%.}
\begin{ruledtabular}
\begin{tabular}{ccccccccccc}
& $s$ & $p_y$ & $p_z$ &  $p_x$ & $d_{xy}$ & $d_{yz}$ & $d_{3z^2-r^2}$ & $d_{xz}$ & $d_{x^2-y^2}$ & Interferences\\
\hline
GR-WBL & 1.3\% & 0.2\% & 0.3\% & 0.1\% & 6.1\% & \textbf{42.4\%} & \textbf{35.8\%} & 17.8\% & 4.7\% & -8.7\% \\
GB-WBL & 13.0\% & 0.1\% & 1.5\% & 0.1\% & 4.7\% & \textbf{19.1\%} & \textbf{83.5\%} & 8.1\% & 3.6\% & -33.7\% \\
GN-WBL & 9.5\% & 0.2\% & 0.3\% & 0.1\% & 3.5\% & \textbf{55.3\%} & 1.7\% & \textbf{23.2\%} & 2.7\% & 3.5\% \\
GR-WSH & 0.8\% & 0.1\% & 0.4\% & 0.0\% & 1.8\% & \textbf{37.5\%} & \textbf{18.6\%} & \textbf{18.6\%} & 7.0\% & 15.2\% \\
GB-WSH & 5.7\% & 0.0\% & 1.7\% & 0.0\% & 1.0\% & \textbf{12.0\%} & \textbf{28.8\%} & 5.7\% & 3.8\% & 41.3\% \\
GN-WSH & 8.4\% & 0.1\% & 0.7\% & 0.0\% & 1.3\% & \textbf{56.8\%} & 1.7\% & \textbf{26.9\%} & 5.1\% & -1.0\% \\
GR-WCA & 0.5\% & \textbf{45.7\%} & \textbf{40.7\%} & 13.2\% & - & - & - & - & - & -0.1\%  \\
GB-WCA & 2.1\% & \textbf{10.2\%} & \textbf{85.1\%} & 2.9\% & - & - & - & - & - & -0.3\% \\
GN-WCA & 2.6\% & \textbf{53.2\%} & \textbf{29.5}\% & 14.9\% & - & - & - & - & - & -0.2\%
\end{tabular}
\end{ruledtabular}
\label{tabp}
\end{table*}

Having a first sight at the colored absolute (Fig.~\ref{fig4}) and relative (Fig.~\ref{fig5}) contributions to the $dI/dV$ it is clear that both the surface and tip properties affect these. One of the most apparent observation is for the row of the WCA tip, where $d$ tip orbitals are absent and $p$ tip orbitals are contributing almost exclusively, which are represented by various shades of red color. Here, the $p_z$ tip orbital has the largest contribution at negative voltages for GR, and in the full voltage range for GB, but for GN only in the vicinity of the earlier mentioned $dI/dV$ peak at about +0.25 V. For the WCA tip, the $p_y$ tip orbitals have generally larger contributions than $p_x$ in the full energy range, and this is due to two effects: (i) $y$ is the direction of the nearest neighbor C atom with respect to the C (GR), B (GB) or N (GN) atomic positions, above which the tip is placed (see the bottom left subfigure in Fig.~\ref{fig2}, where N is placed in the middle of the 3 bright C sites), and (ii) the used anisotropic W(110) tip with $x$ coinciding with the $[1\overline{1}0]$ and $y$ with the $[001]$ tungsten crystallographic direction \cite{Teobaldi12}. Furthermore, for the WCA tip the $s$ tip orbital contributions are generally small, and they are only visible for GN in the vicinity of +0.25 V in Fig.~\ref{fig5}. Tip-orbital interference contributions are negligible for WCA tips on the studied surfaces.

The situation is far more complex when using the tip models with a tungsten apex atom, WBL and WSH, because of the presence of $d$ electron orbitals of these tips, which contribute mostly to the $dI/dV$. Generally, the $p$ tip orbitals lose their importance in these cases, and $s$ tip orbitals contribute visibly for the doped graphene sheets, GB and GN only. Overall, the $s$ tip orbitals have the largest relative contribution in the full bias voltage range for the GB-WBL combination, exceeding 10\%, and for GN we find again that the largest relative $s$ tip orbital contribution is localized around the $dI/dV$ peak at about +0.25 V for both WBL and WSH tips.

The direct contributions of $d$ tip orbitals vary considerably depending on the surface-tip combination. These are fairly constant in energy for GR-WBL, but there is a visible step in the $d$ tip orbital contributions at 0 V for GR-WSH, and their energy-dependent behavior is even more complex in the other cases. The two largest directly contributing tip orbitals are $d_{yz}$ and $d_{3z^2-r^2}$ for GR (with the opposite order for GR-WSH at negative voltages), a dominant $d_{3z^2-r^2}$ and $d_{yz}$ for GB, and a dominant $d_{yz}$ and $d_{xz}$ for GN, but all have an energy dependence as visible in Fig.~\ref{fig5}. Again, the $d_{yz}$ contributions are systematically larger than those of $d_{xz}$ because of the nearest neighbor carbon $y$ direction and the anisotropic shape of the W(110) tip in $x$ and $y$, as discussed above for the relation between $p_y$ and $p_x$. Interestingly, the $d_{3z^2-r^2}$ tip orbital contributions for GN are very small, which is most likely due to the special electronic structure distribution around the N dopant \cite{Telychko14,Telychko15}. Energy-averaged $d_{xy}$ tip orbital contributions are always smaller than 6.1\% (largest for GR-WBL at positive voltages), and those of $d_{x^2-y^2}$ are below 16\% (largest for GN-WSH at negative voltages).

So far, we described the most important findings for the direct tip orbital contributions to the $dI/dV$. Next, let us discuss the effect of the tip orbital interference contributions for the W-apex tips (WBL and WSH), which also exhibit a great variety depending on the surface-tip combination. As can be seen from the Tables \ref{tabf}, \ref{tabm} and \ref{tabp}, in effect, the WBL tip provides negative tip orbital interference contributions for GR and GB, while the WSH tip has positive corresponding contributions. As already mentioned earlier, the GN behaves differently, and the total tip orbital interference contributions change sign with respect to either GR or GB, and they are also smaller. In Fig.~\ref{fig5} the largest tip orbital interference contributions are explicitly shown. The $s+p_z$ contributions are always positive, and they are present in various extents for the studied surface-tip combinations correlating with the direct $s$ tip-orbital contributions, but they are generally small. The $s+d_{3z^2-r^2}$ contributions are much larger in all cases, their value in energy correlates well with both the $s$ and $d_{3z^2-r^2}$ direct contributions, and they can be either positive (for GR-WSH, GB-WSH and GN-WBL) or negative (for GR-WBL, GB-WBL and GN-WSH). The largest negative $s+d_{3z^2-r^2}$ tip orbital interference contributions are found for GB-WBL in the full energy range. Notable $p_y+d_{yz}$ tip orbital interference contributions appear for GR-WBL and GN-WSH, correlating with the direct $d_{yz}$ contributions, and notable $p_z+d_{3z^2-r^2}$ interference contributions are found for GR-WSH and GB-WSH, correlating with those of $d_{3z^2-r^2}$. All other interference contributions are summed up and are denoted by Others in Fig.~\ref{fig5}, which are so small that they are not visible.

The above analysis and discussion imply that the energy-resolved tip-orbital $dI/dV(V)$ contribution maps can be interpreted as fingerprints for the particular surface-tip combinations. Such a simulated $dI/dV$ fingerprint dataset can be further extended by calculating their spatial dependence following a scanning tip at different tunneling parameters and considering a large variety of tip apex structures \cite{Mandi15progsurfsci}, which allows for data analysis using machine learning methods \cite{Guerrero-Rivera24} with the ultimate future goal of realizing predictive STS.

\section{Conclusions}
\label{sec:conc}

We implemented the revised Chen's derivative rule for computationally efficient calculations of the elastic tunneling differential conductance $dI/dV$ in STM junctions based on electronic structures obtained from first principles. The probing tip is included through a single tip apex atom, and its electronic structure can be modeled as a linear combination of electron orbitals of various symmetries, or can be directly transferred from first-principles electronic structure calculations. The $dI/dV$ is calculated in a direct way, and not by performing numerical derivatives of the tunneling current, thus noisy $dI/dV$ simulated data is excluded.

The applicability of the simulation method is demonstrated by performing numerical calculations of the $dI/dV$ on pristine and boron-/nitrogen-doped graphene sheets. As a validation, by comparing our results with those obtained by using an $s$ tip orbital in the spirit of the Tersoff-Hamann approximation, we found an excellent agreement. Furthermore, going beyond the $s$ tip orbital we analyzed the behavior of $dI/dV$ spectra obtained by an electronically featureless $p_z$ tip orbital and a combination of $s+p_z$ tip orbitals. The destructive interference in the latter case results in a reduced conductance in the full energy range, while the local electronic structure of the sample surface affects the $dI/dV$ spectra obtained with a $p_z$ tip in relation to an $s$ tip: enhanced conductance values are found for pristine and B-doped graphene due to the oriented $p_z$ tip, and a systematically reduced conductance appears for N-doped graphene, which is connected to the special spatial distribution of electronic states in the vicinity of the N dopant. Calculation of 2D $dI/dV(x,y)$ (STS) maps also supports the observed difference between B and N dopants in graphene.

We also compared $dI/dV$ spectra at constant-current and constant-height conditions between our implementation and employing Bardeen's tunneling model by taking different tungsten-based probing tips. While the general shape of the differential conductance spectra is qualitatively similar, we discussed the reasons of variations in their fine details between the two simulation methods.

The main advantages of our $dI/dV$ calculation method are: (i) its computational efficiency which increases with a better spatial resolution of calculated STS maps, and (ii) the possibility to decompose the $dI/dV$ in terms of direct and interference tip-orbital contributions. We highlight that the energy-resolved direct and interference contributions to $dI/dV$ arising from the tip's electron orbitals result in a fingerprint of the particular combined surface-tip system, thus indirect information on the dopant's electronic characteristics can also be obtained. For example, for N-doped graphene we found that $d_{3z^2-r^2}$ tip orbitals have very small contributions, unlike for B-doped graphene, which reflects the difference in the spatial localization of their electronic states.

The presented theoretical method allows for an unprecedented analysis of the electron tunneling process in terms of tip-orbital-resolved energy-dependent $dI/dV$ maps, which is not only useful for a deeper physical understanding of particular local surface electronic structure properties of novel material surfaces, but the simulated large amount of data at different tunneling parameters could be treated with machine learning methods for identifying trends in the high-dimensional $dI/dV$ fingerprint dataset at a moderate computational cost. This could lay the basis for realizing predictive STS in the future.

\begin{acknowledgments}
Financial support of the HUN-REN Hungarian Research Network, the National Research Development and Innovation Office of Hungary under Project Nos.\ FK124100 and K138714, the János Bolyai Research Scholarship of the Hungarian Academy of Sciences (Grant No. BO/292/21/11), the Hungarian State
Eötvös Fellowship of the Tempus Public Foundation (Grant No. 2017-11), and a Stipendium Hungaricum Scholarship of the Tempus Public Foundation are gratefully acknowledged.
\end{acknowledgments}


\begin{thebibliography}{55}%
\makeatletter
\providecommand \@ifxundefined [1]{%
 \@ifx{#1\undefined}
}%
\providecommand \@ifnum [1]{%
 \ifnum #1\expandafter \@firstoftwo
 \else \expandafter \@secondoftwo
 \fi
}%
\providecommand \@ifx [1]{%
 \ifx #1\expandafter \@firstoftwo
 \else \expandafter \@secondoftwo
 \fi
}%
\providecommand \natexlab [1]{#1}%
\providecommand \enquote  [1]{``#1''}%
\providecommand \bibnamefont  [1]{#1}%
\providecommand \bibfnamefont [1]{#1}%
\providecommand \citenamefont [1]{#1}%
\providecommand \href@noop [0]{\@secondoftwo}%
\providecommand \href [0]{\begingroup \@sanitize@url \@href}%
\providecommand \@href[1]{\@@startlink{#1}\@@href}%
\providecommand \@@href[1]{\endgroup#1\@@endlink}%
\providecommand \@sanitize@url [0]{\catcode `\\12\catcode `\$12\catcode `\&12\catcode `\#12\catcode `\^12\catcode `\_12\catcode `\%12\relax}%
\providecommand \@@startlink[1]{}%
\providecommand \@@endlink[0]{}%
\providecommand \url  [0]{\begingroup\@sanitize@url \@url }%
\providecommand \@url [1]{\endgroup\@href {#1}{\urlprefix }}%
\providecommand \urlprefix  [0]{URL }%
\providecommand \Eprint [0]{\href }%
\providecommand \doibase [0]{https://doi.org/}%
\providecommand \selectlanguage [0]{\@gobble}%
\providecommand \bibinfo  [0]{\@secondoftwo}%
\providecommand \bibfield  [0]{\@secondoftwo}%
\providecommand \translation [1]{[#1]}%
\providecommand \BibitemOpen [0]{}%
\providecommand \bibitemStop [0]{}%
\providecommand \bibitemNoStop [0]{.\EOS\space}%
\providecommand \EOS [0]{\spacefactor3000\relax}%
\providecommand \BibitemShut  [1]{\csname bibitem#1\endcsname}%
\let\auto@bib@innerbib\@empty
\bibitem [{\citenamefont {Binnig}\ \emph {et~al.}(1982)\citenamefont {Binnig}, \citenamefont {Rohrer}, \citenamefont {Gerber},\ and\ \citenamefont {Weibel}}]{Binnig82}%
  \BibitemOpen
  \bibfield  {author} {\bibinfo {author} {\bibfnamefont {G.}~\bibnamefont {Binnig}}, \bibinfo {author} {\bibfnamefont {H.}~\bibnamefont {Rohrer}}, \bibinfo {author} {\bibfnamefont {C.}~\bibnamefont {Gerber}},\ and\ \bibinfo {author} {\bibfnamefont {E.}~\bibnamefont {Weibel}},\ }\bibfield  {title} {\bibinfo {title} {Surface studies by scanning tunneling microscopy},\ }\href {https://doi.org/10.1103/PhysRevLett.49.57} {\bibfield  {journal} {\bibinfo  {journal} {Phys. Rev. Lett.}\ }\textbf {\bibinfo {volume} {49}},\ \bibinfo {pages} {57} (\bibinfo {year} {1982})}\BibitemShut {NoStop}%
\bibitem [{\citenamefont {Palot{\'a}s}\ \emph {et~al.}(2014)\citenamefont {Palot{\'a}s}, \citenamefont {M{\'a}ndi},\ and\ \citenamefont {Hofer}}]{Palotas14}%
  \BibitemOpen
  \bibfield  {author} {\bibinfo {author} {\bibfnamefont {K.}~\bibnamefont {Palot{\'a}s}}, \bibinfo {author} {\bibfnamefont {G.}~\bibnamefont {M{\'a}ndi}},\ and\ \bibinfo {author} {\bibfnamefont {W.~A.}\ \bibnamefont {Hofer}},\ }\bibfield  {title} {\bibinfo {title} {{Three-dimensional Wentzel-Kramers-Brillouin approach for the simulation of scanning tunneling microscopy and spectroscopy}},\ }\href {https://doi.org/10.1007/s11467-013-0354-4} {\bibfield  {journal} {\bibinfo  {journal} {Frontiers of Physics}\ }\textbf {\bibinfo {volume} {9}},\ \bibinfo {pages} {711} (\bibinfo {year} {2014})}\BibitemShut {NoStop}%
\bibitem [{\citenamefont {Tersoff}\ and\ \citenamefont {Hamann}(1983)}]{Tersoff83}%
  \BibitemOpen
  \bibfield  {author} {\bibinfo {author} {\bibfnamefont {J.}~\bibnamefont {Tersoff}}\ and\ \bibinfo {author} {\bibfnamefont {D.~R.}\ \bibnamefont {Hamann}},\ }\bibfield  {title} {\bibinfo {title} {Theory and application for the scanning tunneling microscope},\ }\href {https://doi.org/10.1103/PhysRevLett.50.1998} {\bibfield  {journal} {\bibinfo  {journal} {Phys. Rev. Lett.}\ }\textbf {\bibinfo {volume} {50}},\ \bibinfo {pages} {1998} (\bibinfo {year} {1983})}\BibitemShut {NoStop}%
\bibitem [{\citenamefont {Tersoff}\ and\ \citenamefont {Hamann}(1985)}]{Tersoff85}%
  \BibitemOpen
  \bibfield  {author} {\bibinfo {author} {\bibfnamefont {J.}~\bibnamefont {Tersoff}}\ and\ \bibinfo {author} {\bibfnamefont {D.~R.}\ \bibnamefont {Hamann}},\ }\bibfield  {title} {\bibinfo {title} {Theory of the scanning tunneling microscope},\ }\href {https://doi.org/10.1103/PhysRevB.31.805} {\bibfield  {journal} {\bibinfo  {journal} {Phys. Rev. B}\ }\textbf {\bibinfo {volume} {31}},\ \bibinfo {pages} {805} (\bibinfo {year} {1985})}\BibitemShut {NoStop}%
\bibitem [{\citenamefont {Chen}(1990)}]{Chen90}%
  \BibitemOpen
  \bibfield  {author} {\bibinfo {author} {\bibfnamefont {C.~J.}\ \bibnamefont {Chen}},\ }\bibfield  {title} {\bibinfo {title} {Tunneling matrix elements in three-dimensional space: The derivative rule and the sum rule},\ }\href {https://doi.org/10.1103/PhysRevB.42.8841} {\bibfield  {journal} {\bibinfo  {journal} {Phys. Rev. B}\ }\textbf {\bibinfo {volume} {42}},\ \bibinfo {pages} {8841} (\bibinfo {year} {1990})}\BibitemShut {NoStop}%
\bibitem [{\citenamefont {M\'andi}\ and\ \citenamefont {Palot\'as}(2015)}]{Mandi15}%
  \BibitemOpen
  \bibfield  {author} {\bibinfo {author} {\bibfnamefont {G.}~\bibnamefont {M\'andi}}\ and\ \bibinfo {author} {\bibfnamefont {K.}~\bibnamefont {Palot\'as}},\ }\bibfield  {title} {\bibinfo {title} {{Chen's derivative rule revisited: Role of tip-orbital interference in STM}},\ }\href {https://doi.org/10.1103/PhysRevB.91.165406} {\bibfield  {journal} {\bibinfo  {journal} {Phys. Rev. B}\ }\textbf {\bibinfo {volume} {91}},\ \bibinfo {pages} {165406} (\bibinfo {year} {2015})}\BibitemShut {NoStop}%
\bibitem [{\citenamefont {Bardeen}(1961)}]{Bardeen61}%
  \BibitemOpen
  \bibfield  {author} {\bibinfo {author} {\bibfnamefont {J.}~\bibnamefont {Bardeen}},\ }\bibfield  {title} {\bibinfo {title} {Tunnelling from a many-particle point of view},\ }\href {https://doi.org/10.1103/PhysRevLett.6.57} {\bibfield  {journal} {\bibinfo  {journal} {Phys. Rev. Lett.}\ }\textbf {\bibinfo {volume} {6}},\ \bibinfo {pages} {57} (\bibinfo {year} {1961})}\BibitemShut {NoStop}%
\bibitem [{\citenamefont {Blanco}\ \emph {et~al.}(2004)\citenamefont {Blanco}, \citenamefont {Gonz\'alez}, \citenamefont {Jel\'{\i}nek}, \citenamefont {Ortega}, \citenamefont {Flores},\ and\ \citenamefont {P\'erez}}]{Blanco04}%
  \BibitemOpen
  \bibfield  {author} {\bibinfo {author} {\bibfnamefont {J.~M.}\ \bibnamefont {Blanco}}, \bibinfo {author} {\bibfnamefont {C.}~\bibnamefont {Gonz\'alez}}, \bibinfo {author} {\bibfnamefont {P.}~\bibnamefont {Jel\'{\i}nek}}, \bibinfo {author} {\bibfnamefont {J.}~\bibnamefont {Ortega}}, \bibinfo {author} {\bibfnamefont {F.}~\bibnamefont {Flores}},\ and\ \bibinfo {author} {\bibfnamefont {R.}~\bibnamefont {P\'erez}},\ }\bibfield  {title} {\bibinfo {title} {{First-principles simulations of STM images: From tunneling to the contact regime}},\ }\href {https://doi.org/10.1103/PhysRevB.70.085405} {\bibfield  {journal} {\bibinfo  {journal} {Phys. Rev. B}\ }\textbf {\bibinfo {volume} {70}},\ \bibinfo {pages} {085405} (\bibinfo {year} {2004})}\BibitemShut {NoStop}%
\bibitem [{\citenamefont {Blanco}\ \emph {et~al.}(2006)\citenamefont {Blanco}, \citenamefont {Flores},\ and\ \citenamefont {Pérez}}]{Blanco06}%
  \BibitemOpen
  \bibfield  {author} {\bibinfo {author} {\bibfnamefont {J.~M.}\ \bibnamefont {Blanco}}, \bibinfo {author} {\bibfnamefont {F.}~\bibnamefont {Flores}},\ and\ \bibinfo {author} {\bibfnamefont {R.}~\bibnamefont {Pérez}},\ }\bibfield  {title} {\bibinfo {title} {{STM-theory: Image potential, chemistry and surface relaxation}},\ }\href {https://doi.org/https://doi.org/10.1016/j.progsurf.2006.07.004} {\bibfield  {journal} {\bibinfo  {journal} {Prog. Surf. Sci.}\ }\textbf {\bibinfo {volume} {81}},\ \bibinfo {pages} {403} (\bibinfo {year} {2006})}\BibitemShut {NoStop}%
\bibitem [{\citenamefont {González}\ \emph {et~al.}(2016)\citenamefont {González}, \citenamefont {Abad}, \citenamefont {Dappe},\ and\ \citenamefont {Cuevas}}]{González16}%
  \BibitemOpen
  \bibfield  {author} {\bibinfo {author} {\bibfnamefont {C.}~\bibnamefont {González}}, \bibinfo {author} {\bibfnamefont {E.}~\bibnamefont {Abad}}, \bibinfo {author} {\bibfnamefont {Y.~J.}\ \bibnamefont {Dappe}},\ and\ \bibinfo {author} {\bibfnamefont {J.~C.}\ \bibnamefont {Cuevas}},\ }\bibfield  {title} {\bibinfo {title} {Theoretical study of carbon-based tips for scanning tunnelling microscopy},\ }\href {https://doi.org/10.1088/0957-4484/27/10/105201} {\bibfield  {journal} {\bibinfo  {journal} {Nanotechnology}\ }\textbf {\bibinfo {volume} {27}},\ \bibinfo {pages} {105201} (\bibinfo {year} {2016})}\BibitemShut {NoStop}%
\bibitem [{\citenamefont {Lang}(1986)}]{Lang86}%
  \BibitemOpen
  \bibfield  {author} {\bibinfo {author} {\bibfnamefont {N.~D.}\ \bibnamefont {Lang}},\ }\bibfield  {title} {\bibinfo {title} {Spectroscopy of single atoms in the scanning tunneling microscope},\ }\href {https://doi.org/10.1103/PhysRevB.34.5947} {\bibfield  {journal} {\bibinfo  {journal} {Phys. Rev. B}\ }\textbf {\bibinfo {volume} {34}},\ \bibinfo {pages} {5947} (\bibinfo {year} {1986})}\BibitemShut {NoStop}%
\bibitem [{\citenamefont {Ukraintsev}(1996)}]{PhysRevB.53.11176}%
  \BibitemOpen
  \bibfield  {author} {\bibinfo {author} {\bibfnamefont {V.~A.}\ \bibnamefont {Ukraintsev}},\ }\bibfield  {title} {\bibinfo {title} {Data evaluation technique for electron-tunneling spectroscopy},\ }\href {https://doi.org/10.1103/PhysRevB.53.11176} {\bibfield  {journal} {\bibinfo  {journal} {Phys. Rev. B}\ }\textbf {\bibinfo {volume} {53}},\ \bibinfo {pages} {11176} (\bibinfo {year} {1996})}\BibitemShut {NoStop}%
\bibitem [{\citenamefont {Koslowski}\ \emph {et~al.}(2007)\citenamefont {Koslowski}, \citenamefont {Dietrich}, \citenamefont {Tschetschetkin},\ and\ \citenamefont {Ziemann}}]{PhysRevB.75.035421}%
  \BibitemOpen
  \bibfield  {author} {\bibinfo {author} {\bibfnamefont {B.}~\bibnamefont {Koslowski}}, \bibinfo {author} {\bibfnamefont {C.}~\bibnamefont {Dietrich}}, \bibinfo {author} {\bibfnamefont {A.}~\bibnamefont {Tschetschetkin}},\ and\ \bibinfo {author} {\bibfnamefont {P.}~\bibnamefont {Ziemann}},\ }\bibfield  {title} {\bibinfo {title} {Evaluation of scanning tunneling spectroscopy data: Approaching a quantitative determination of the electronic density of states},\ }\href {https://doi.org/10.1103/PhysRevB.75.035421} {\bibfield  {journal} {\bibinfo  {journal} {Phys. Rev. B}\ }\textbf {\bibinfo {volume} {75}},\ \bibinfo {pages} {035421} (\bibinfo {year} {2007})}\BibitemShut {NoStop}%
\bibitem [{\citenamefont {Passoni}\ \emph {et~al.}(2009)\citenamefont {Passoni}, \citenamefont {Donati}, \citenamefont {Li~Bassi}, \citenamefont {Casari},\ and\ \citenamefont {Bottani}}]{Passoni09}%
  \BibitemOpen
  \bibfield  {author} {\bibinfo {author} {\bibfnamefont {M.}~\bibnamefont {Passoni}}, \bibinfo {author} {\bibfnamefont {F.}~\bibnamefont {Donati}}, \bibinfo {author} {\bibfnamefont {A.}~\bibnamefont {Li~Bassi}}, \bibinfo {author} {\bibfnamefont {C.~S.}\ \bibnamefont {Casari}},\ and\ \bibinfo {author} {\bibfnamefont {C.~E.}\ \bibnamefont {Bottani}},\ }\bibfield  {title} {\bibinfo {title} {Recovery of local density of states using scanning tunneling spectroscopy},\ }\href {https://doi.org/10.1103/PhysRevB.79.045404} {\bibfield  {journal} {\bibinfo  {journal} {Phys. Rev. B}\ }\textbf {\bibinfo {volume} {79}},\ \bibinfo {pages} {045404} (\bibinfo {year} {2009})}\BibitemShut {NoStop}%
\bibitem [{\citenamefont {Ziegler}\ \emph {et~al.}(2009)\citenamefont {Ziegler}, \citenamefont {N\'eel}, \citenamefont {Sperl}, \citenamefont {Kr\"oger},\ and\ \citenamefont {Berndt}}]{PhysRevB.80.125402}%
  \BibitemOpen
  \bibfield  {author} {\bibinfo {author} {\bibfnamefont {M.}~\bibnamefont {Ziegler}}, \bibinfo {author} {\bibfnamefont {N.}~\bibnamefont {N\'eel}}, \bibinfo {author} {\bibfnamefont {A.}~\bibnamefont {Sperl}}, \bibinfo {author} {\bibfnamefont {J.}~\bibnamefont {Kr\"oger}},\ and\ \bibinfo {author} {\bibfnamefont {R.}~\bibnamefont {Berndt}},\ }\bibfield  {title} {\bibinfo {title} {Local density of states from constant-current tunneling spectra},\ }\href {https://doi.org/10.1103/PhysRevB.80.125402} {\bibfield  {journal} {\bibinfo  {journal} {Phys. Rev. B}\ }\textbf {\bibinfo {volume} {80}},\ \bibinfo {pages} {125402} (\bibinfo {year} {2009})}\BibitemShut {NoStop}%
\bibitem [{\citenamefont {Koslowski}\ \emph {et~al.}(2009)\citenamefont {Koslowski}, \citenamefont {Pfeifer},\ and\ \citenamefont {Ziemann}}]{PhysRevB.80.165419}%
  \BibitemOpen
  \bibfield  {author} {\bibinfo {author} {\bibfnamefont {B.}~\bibnamefont {Koslowski}}, \bibinfo {author} {\bibfnamefont {H.}~\bibnamefont {Pfeifer}},\ and\ \bibinfo {author} {\bibfnamefont {P.}~\bibnamefont {Ziemann}},\ }\bibfield  {title} {\bibinfo {title} {Deconvolution of the electronic density of states of tip and sample from scanning tunneling spectroscopy data: Proof of principle},\ }\href {https://doi.org/10.1103/PhysRevB.80.165419} {\bibfield  {journal} {\bibinfo  {journal} {Phys. Rev. B}\ }\textbf {\bibinfo {volume} {80}},\ \bibinfo {pages} {165419} (\bibinfo {year} {2009})}\BibitemShut {NoStop}%
\bibitem [{\citenamefont {Hofer}\ and\ \citenamefont {Garcia-Lekue}(2005)}]{Hofer05}%
  \BibitemOpen
  \bibfield  {author} {\bibinfo {author} {\bibfnamefont {W.~A.}\ \bibnamefont {Hofer}}\ and\ \bibinfo {author} {\bibfnamefont {A.}~\bibnamefont {Garcia-Lekue}},\ }\bibfield  {title} {\bibinfo {title} {Differential tunneling spectroscopy simulations: Imaging surface states},\ }\href {https://doi.org/10.1103/PhysRevB.71.085401} {\bibfield  {journal} {\bibinfo  {journal} {Phys. Rev. B}\ }\textbf {\bibinfo {volume} {71}},\ \bibinfo {pages} {085401} (\bibinfo {year} {2005})}\BibitemShut {NoStop}%
\bibitem [{\citenamefont {Palot\'as}\ \emph {et~al.}(2011{\natexlab{a}})\citenamefont {Palot\'as}, \citenamefont {Hofer},\ and\ \citenamefont {Szunyogh}}]{Palotas11}%
  \BibitemOpen
  \bibfield  {author} {\bibinfo {author} {\bibfnamefont {K.}~\bibnamefont {Palot\'as}}, \bibinfo {author} {\bibfnamefont {W.~A.}\ \bibnamefont {Hofer}},\ and\ \bibinfo {author} {\bibfnamefont {L.}~\bibnamefont {Szunyogh}},\ }\bibfield  {title} {\bibinfo {title} {{Simulation of spin-polarized scanning tunneling microscopy on complex magnetic surfaces: Case of a Cr monolayer on Ag(111)}},\ }\href {https://doi.org/10.1103/PhysRevB.84.174428} {\bibfield  {journal} {\bibinfo  {journal} {Phys. Rev. B}\ }\textbf {\bibinfo {volume} {84}},\ \bibinfo {pages} {174428} (\bibinfo {year} {2011}{\natexlab{a}})}\BibitemShut {NoStop}%
\bibitem [{\citenamefont {Palot\'as}\ \emph {et~al.}(2011{\natexlab{b}})\citenamefont {Palot\'as}, \citenamefont {Hofer},\ and\ \citenamefont {Szunyogh}}]{Palotas11sts}%
  \BibitemOpen
  \bibfield  {author} {\bibinfo {author} {\bibfnamefont {K.}~\bibnamefont {Palot\'as}}, \bibinfo {author} {\bibfnamefont {W.~A.}\ \bibnamefont {Hofer}},\ and\ \bibinfo {author} {\bibfnamefont {L.}~\bibnamefont {Szunyogh}},\ }\bibfield  {title} {\bibinfo {title} {Theoretical study of the role of the tip in enhancing the sensitivity of differential conductance tunneling spectroscopy on magnetic surfaces},\ }\href {https://doi.org/10.1103/PhysRevB.83.214410} {\bibfield  {journal} {\bibinfo  {journal} {Phys. Rev. B}\ }\textbf {\bibinfo {volume} {83}},\ \bibinfo {pages} {214410} (\bibinfo {year} {2011}{\natexlab{b}})}\BibitemShut {NoStop}%
\bibitem [{\citenamefont {Palot\'as}\ \emph {et~al.}(2012)\citenamefont {Palot\'as}, \citenamefont {Hofer},\ and\ \citenamefont {Szunyogh}}]{Palotas12}%
  \BibitemOpen
  \bibfield  {author} {\bibinfo {author} {\bibfnamefont {K.}~\bibnamefont {Palot\'as}}, \bibinfo {author} {\bibfnamefont {W.~A.}\ \bibnamefont {Hofer}},\ and\ \bibinfo {author} {\bibfnamefont {L.}~\bibnamefont {Szunyogh}},\ }\bibfield  {title} {\bibinfo {title} {{Simulation of spin-polarized scanning tunneling spectroscopy on complex magnetic surfaces: Case of a Cr monolayer on Ag(111)}},\ }\href {https://doi.org/10.1103/PhysRevB.85.205427} {\bibfield  {journal} {\bibinfo  {journal} {Phys. Rev. B}\ }\textbf {\bibinfo {volume} {85}},\ \bibinfo {pages} {205427} (\bibinfo {year} {2012})}\BibitemShut {NoStop}%
\bibitem [{\citenamefont {Palot\'as}(2013)}]{Palotas13}%
  \BibitemOpen
  \bibfield  {author} {\bibinfo {author} {\bibfnamefont {K.}~\bibnamefont {Palot\'as}},\ }\bibfield  {title} {\bibinfo {title} {Prediction of the bias voltage dependent magnetic contrast in spin-polarized scanning tunneling microscopy},\ }\href {https://doi.org/10.1103/PhysRevB.87.024417} {\bibfield  {journal} {\bibinfo  {journal} {Phys. Rev. B}\ }\textbf {\bibinfo {volume} {87}},\ \bibinfo {pages} {024417} (\bibinfo {year} {2013})}\BibitemShut {NoStop}%
\bibitem [{\citenamefont {Jurczyszyn}\ and\ \citenamefont {Stankiewicz}(2003)}]{Jurczyszyn03}%
  \BibitemOpen
  \bibfield  {author} {\bibinfo {author} {\bibfnamefont {L.}~\bibnamefont {Jurczyszyn}}\ and\ \bibinfo {author} {\bibfnamefont {B.}~\bibnamefont {Stankiewicz}},\ }\bibfield  {title} {\bibinfo {title} {{Inter-orbital interference in STM tip during electron tunneling in tip–sample system: influence on STM images}},\ }\href {https://doi.org/https://doi.org/10.1016/j.progsurf.2003.08.014} {\bibfield  {journal} {\bibinfo  {journal} {Prog. Surf. Sci.}\ }\textbf {\bibinfo {volume} {74}},\ \bibinfo {pages} {185} (\bibinfo {year} {2003})}\BibitemShut {NoStop}%
\bibitem [{\citenamefont {Jurczyszyn}\ and\ \citenamefont {Stankiewicz}(2005)}]{Jurczyszyn05}%
  \BibitemOpen
  \bibfield  {author} {\bibinfo {author} {\bibfnamefont {L.}~\bibnamefont {Jurczyszyn}}\ and\ \bibinfo {author} {\bibfnamefont {B.}~\bibnamefont {Stankiewicz}},\ }\bibfield  {title} {\bibinfo {title} {{The role of interorbital interference in the formation of STS spectra}},\ }\href {https://doi.org/https://doi.org/10.1016/j.apsusc.2004.07.066} {\bibfield  {journal} {\bibinfo  {journal} {Appl. Surf. Sci.}\ }\textbf {\bibinfo {volume} {242}},\ \bibinfo {pages} {70} (\bibinfo {year} {2005})}\BibitemShut {NoStop}%
\bibitem [{\citenamefont {S\l{}awi\ifmmode~\acute{n}\else \'{n}\fi{}ska}\ and\ \citenamefont {Zasada}(2011)}]{Slawinska11}%
  \BibitemOpen
  \bibfield  {author} {\bibinfo {author} {\bibfnamefont {J.}~\bibnamefont {S\l{}awi\ifmmode~\acute{n}\else \'{n}\fi{}ska}}\ and\ \bibinfo {author} {\bibfnamefont {I.}~\bibnamefont {Zasada}},\ }\bibfield  {title} {\bibinfo {title} {{Fingerprints of Dirac points in first-principles calculations of scanning tunneling spectra of graphene on a metal substrate}},\ }\href {https://doi.org/10.1103/PhysRevB.84.235445} {\bibfield  {journal} {\bibinfo  {journal} {Phys. Rev. B}\ }\textbf {\bibinfo {volume} {84}},\ \bibinfo {pages} {235445} (\bibinfo {year} {2011})}\BibitemShut {NoStop}%
\bibitem [{\citenamefont {Krej\ifmmode~\check{c}\else \v{c}\fi{}\'{\i}}\ \emph {et~al.}(2017)\citenamefont {Krej\ifmmode~\check{c}\else \v{c}\fi{}\'{\i}}, \citenamefont {Hapala}, \citenamefont {Ondr\'a\ifmmode~\check{c}\else \v{c}\fi{}ek},\ and\ \citenamefont {Jel\'{\i}nek}}]{Krejci17}%
  \BibitemOpen
  \bibfield  {author} {\bibinfo {author} {\bibfnamefont {O.}~\bibnamefont {Krej\ifmmode~\check{c}\else \v{c}\fi{}\'{\i}}}, \bibinfo {author} {\bibfnamefont {P.}~\bibnamefont {Hapala}}, \bibinfo {author} {\bibfnamefont {M.}~\bibnamefont {Ondr\'a\ifmmode~\check{c}\else \v{c}\fi{}ek}},\ and\ \bibinfo {author} {\bibfnamefont {P.}~\bibnamefont {Jel\'{\i}nek}},\ }\bibfield  {title} {\bibinfo {title} {{Principles and simulations of high-resolution STM imaging with a flexible tip apex}},\ }\href {https://doi.org/10.1103/PhysRevB.95.045407} {\bibfield  {journal} {\bibinfo  {journal} {Phys. Rev. B}\ }\textbf {\bibinfo {volume} {95}},\ \bibinfo {pages} {045407} (\bibinfo {year} {2017})}\BibitemShut {NoStop}%
\bibitem [{\citenamefont {Oinonen}\ \emph {et~al.}(2024)\citenamefont {Oinonen}, \citenamefont {Yakutovich}, \citenamefont {Gallardo}, \citenamefont {Ondráček}, \citenamefont {Hapala},\ and\ \citenamefont {Krejčí}}]{Oinonen24}%
  \BibitemOpen
  \bibfield  {author} {\bibinfo {author} {\bibfnamefont {N.}~\bibnamefont {Oinonen}}, \bibinfo {author} {\bibfnamefont {A.~V.}\ \bibnamefont {Yakutovich}}, \bibinfo {author} {\bibfnamefont {A.}~\bibnamefont {Gallardo}}, \bibinfo {author} {\bibfnamefont {M.}~\bibnamefont {Ondráček}}, \bibinfo {author} {\bibfnamefont {P.}~\bibnamefont {Hapala}},\ and\ \bibinfo {author} {\bibfnamefont {O.}~\bibnamefont {Krejčí}},\ }\bibfield  {title} {\bibinfo {title} {Advancing scanning probe microscopy simulations: A decade of development in probe-particle models},\ }\href {https://doi.org/https://doi.org/10.1016/j.cpc.2024.109341} {\bibfield  {journal} {\bibinfo  {journal} {Comput. Phys. Commun.}\ }\textbf {\bibinfo {volume} {305}},\ \bibinfo {pages} {109341} (\bibinfo {year} {2024})}\BibitemShut {NoStop}%
\bibitem [{\citenamefont {Abilio}\ \emph {et~al.}(2024)\citenamefont {Abilio}, \citenamefont {N\'eel}, \citenamefont {Kr\"oger},\ and\ \citenamefont {Palot\'as}}]{Abilio24}%
  \BibitemOpen
  \bibfield  {author} {\bibinfo {author} {\bibfnamefont {I.}~\bibnamefont {Abilio}}, \bibinfo {author} {\bibfnamefont {N.}~\bibnamefont {N\'eel}}, \bibinfo {author} {\bibfnamefont {J.}~\bibnamefont {Kr\"oger}},\ and\ \bibinfo {author} {\bibfnamefont {K.}~\bibnamefont {Palot\'as}},\ }\bibfield  {title} {\bibinfo {title} {{Scanning tunneling microscopy using CO-terminated probes with tilted and straight geometries}},\ }\href {https://doi.org/10.1103/PhysRevB.110.125422} {\bibfield  {journal} {\bibinfo  {journal} {Phys. Rev. B}\ }\textbf {\bibinfo {volume} {110}},\ \bibinfo {pages} {125422} (\bibinfo {year} {2024})}\BibitemShut {NoStop}%
\bibitem [{\citenamefont {Setescak}\ \emph {et~al.}(2025)\citenamefont {Setescak}, \citenamefont {Aguilera}, \citenamefont {Weindl}, \citenamefont {Kronseder}, \citenamefont {Donarini},\ and\ \citenamefont {Giessibl}}]{setescak24}%
  \BibitemOpen
  \bibfield  {author} {\bibinfo {author} {\bibfnamefont {C.~S.}\ \bibnamefont {Setescak}}, \bibinfo {author} {\bibfnamefont {I.}~\bibnamefont {Aguilera}}, \bibinfo {author} {\bibfnamefont {A.}~\bibnamefont {Weindl}}, \bibinfo {author} {\bibfnamefont {M.}~\bibnamefont {Kronseder}}, \bibinfo {author} {\bibfnamefont {A.}~\bibnamefont {Donarini}},\ and\ \bibinfo {author} {\bibfnamefont {F.~J.}\ \bibnamefont {Giessibl}},\ }\bibfield  {title} {\bibinfo {title} {{Probing the electronic structure at the boundary of topological insulators in the ${\mathrm{Bi}}_{2}{\mathrm{Se}}_{3}$ family by combined scanning tunneling and atomic force microscopy}},\ }\href {https://doi.org/10.1103/PhysRevB.111.165305} {\bibfield  {journal} {\bibinfo  {journal} {Phys. Rev. B}\ }\textbf {\bibinfo {volume} {111}},\ \bibinfo {pages} {165305} (\bibinfo {year} {2025})}\BibitemShut {NoStop}%
\bibitem [{\citenamefont {Paschke}\ \emph {et~al.}(2025)\citenamefont {Paschke}, \citenamefont {Lieske}, \citenamefont {Albrecht}, \citenamefont {Chen}, \citenamefont {Repp},\ and\ \citenamefont {Gross}}]{Paschke25}%
  \BibitemOpen
  \bibfield  {author} {\bibinfo {author} {\bibfnamefont {F.}~\bibnamefont {Paschke}}, \bibinfo {author} {\bibfnamefont {L.-A.}\ \bibnamefont {Lieske}}, \bibinfo {author} {\bibfnamefont {F.}~\bibnamefont {Albrecht}}, \bibinfo {author} {\bibfnamefont {C.~J.}\ \bibnamefont {Chen}}, \bibinfo {author} {\bibfnamefont {J.}~\bibnamefont {Repp}},\ and\ \bibinfo {author} {\bibfnamefont {L.}~\bibnamefont {Gross}},\ }\bibfield  {title} {\bibinfo {title} {{Distance and voltage dependence of orbital density imaging using a CO-functionalized tip in scanning tunneling microscopy}},\ }\href {https://doi.org/10.1021/acsnano.4c14476} {\bibfield  {journal} {\bibinfo  {journal} {ACS Nano}\ }\textbf {\bibinfo {volume} {19}},\ \bibinfo {pages} {2641} (\bibinfo {year} {2025})}\BibitemShut {NoStop}%
\bibitem [{\citenamefont {Robles}\ \emph {et~al.}(2025)\citenamefont {Robles}, \citenamefont {Li}, \citenamefont {Realista}, \citenamefont {Martinho}, \citenamefont {Gruber}, \citenamefont {Weismann}, \citenamefont {Lorente},\ and\ \citenamefont {Berndt}}]{Robles25}%
  \BibitemOpen
  \bibfield  {author} {\bibinfo {author} {\bibfnamefont {R.}~\bibnamefont {Robles}}, \bibinfo {author} {\bibfnamefont {C.}~\bibnamefont {Li}}, \bibinfo {author} {\bibfnamefont {S.}~\bibnamefont {Realista}}, \bibinfo {author} {\bibfnamefont {P.~N.}\ \bibnamefont {Martinho}}, \bibinfo {author} {\bibfnamefont {M.}~\bibnamefont {Gruber}}, \bibinfo {author} {\bibfnamefont {A.}~\bibnamefont {Weismann}}, \bibinfo {author} {\bibfnamefont {N.}~\bibnamefont {Lorente}},\ and\ \bibinfo {author} {\bibfnamefont {R.}~\bibnamefont {Berndt}},\ }\bibfield  {title} {\bibinfo {title} {{Interpreting tunneling spectroscopic maps of a dinuclear Co(II) complex on gold}},\ }\href {https://doi.org/10.1103/PhysRevB.111.085409} {\bibfield  {journal} {\bibinfo  {journal} {Phys. Rev. B}\ }\textbf {\bibinfo {volume} {111}},\ \bibinfo {pages} {085409} (\bibinfo {year} {2025})}\BibitemShut {NoStop}%
\bibitem [{\citenamefont {Le~Ster}\ \emph {et~al.}(2024)\citenamefont {Le~Ster}, \citenamefont {Dabrowski}, \citenamefont {Krukowski}, \citenamefont {Rogala}, \citenamefont {Lutsyk}, \citenamefont {Ry\ifmmode~\acute{s}\else \'{s}\fi{}}, \citenamefont {M\"arkl}, \citenamefont {Brown}, \citenamefont {S\l{}awi\ifmmode~\acute{n}\else \'{n}\fi{}ska}, \citenamefont {Palot\'as},\ and\ \citenamefont {Kowalczyk}}]{LeSter24}%
  \BibitemOpen
  \bibfield  {author} {\bibinfo {author} {\bibfnamefont {M.}~\bibnamefont {Le~Ster}}, \bibinfo {author} {\bibfnamefont {P.}~\bibnamefont {Dabrowski}}, \bibinfo {author} {\bibfnamefont {P.}~\bibnamefont {Krukowski}}, \bibinfo {author} {\bibfnamefont {M.}~\bibnamefont {Rogala}}, \bibinfo {author} {\bibfnamefont {I.}~\bibnamefont {Lutsyk}}, \bibinfo {author} {\bibfnamefont {W.}~\bibnamefont {Ry\ifmmode~\acute{s}\else \'{s}\fi{}}}, \bibinfo {author} {\bibfnamefont {T.}~\bibnamefont {M\"arkl}}, \bibinfo {author} {\bibfnamefont {S.~A.}\ \bibnamefont {Brown}}, \bibinfo {author} {\bibfnamefont {J.}~\bibnamefont {S\l{}awi\ifmmode~\acute{n}\else \'{n}\fi{}ska}}, \bibinfo {author} {\bibfnamefont {K.}~\bibnamefont {Palot\'as}},\ and\ \bibinfo {author} {\bibfnamefont {P.~J.}\ \bibnamefont {Kowalczyk}},\ }\bibfield  {title} {\bibinfo {title} {Moir\'e plane wave expansion model for scanning tunneling microscopy simulations of incommensurate two-dimensional materials},\ }\href {https://doi.org/10.1103/PhysRevB.110.195418}
  {\bibfield  {journal} {\bibinfo  {journal} {Phys. Rev. B}\ }\textbf {\bibinfo {volume} {110}},\ \bibinfo {pages} {195418} (\bibinfo {year} {2024})}\BibitemShut {NoStop}%
\bibitem [{\citenamefont {Gutiérrez}\ and\ \citenamefont {Prado}(2024)}]{gutierrez24}%
  \BibitemOpen
  \bibfield  {author} {\bibinfo {author} {\bibfnamefont {C.}~\bibnamefont {Gutiérrez}}\ and\ \bibinfo {author} {\bibfnamefont {A.~G.}\ \bibnamefont {Prado}},\ }\href {https://arxiv.org/abs/2412.18332} {\bibinfo {title} {{PyAtoms: An interactive tool for rapidly simulating atomic scanning tunneling microscopy images of 2D materials, moir\'e systems and superlattices}}} (\bibinfo {year} {2024}),\ \Eprint {https://arxiv.org/abs/2412.18332} {arXiv:2412.18332 [cond-mat.mes-hall]} \BibitemShut {NoStop}%
\bibitem [{\citenamefont {Hofer}\ and\ \citenamefont {Redinger}(2000)}]{Hofer00}%
  \BibitemOpen
  \bibfield  {author} {\bibinfo {author} {\bibfnamefont {W.}~\bibnamefont {Hofer}}\ and\ \bibinfo {author} {\bibfnamefont {J.}~\bibnamefont {Redinger}},\ }\bibfield  {title} {\bibinfo {title} {{Scanning tunneling microscopy of binary alloys: first principles calculation of the current for PtX (100) surfaces.}},\ }\href {https://doi.org/10.1016/s0039-6028(99)01053-5} {\bibfield  {journal} {\bibinfo  {journal} {Surf. Sci.}\ }\textbf {\bibinfo {volume} {447}},\ \bibinfo {pages} {51} (\bibinfo {year} {2000})}\BibitemShut {NoStop}%
\bibitem [{\citenamefont {Hofer}(2003)}]{Hofer03}%
  \BibitemOpen
  \bibfield  {author} {\bibinfo {author} {\bibfnamefont {W.}~\bibnamefont {Hofer}},\ }\bibfield  {title} {\bibinfo {title} {Challenges and errors: Interpreting high resolution images in scanning tunneling microscopy},\ }\href {https://doi.org/10.1016/S0079-6816(03)00005-4} {\bibfield  {journal} {\bibinfo  {journal} {Prog. Surf. Sci.}\ }\textbf {\bibinfo {volume} {71}},\ \bibinfo {pages} {147} (\bibinfo {year} {2003})}\BibitemShut {NoStop}%
\bibitem [{\citenamefont {Palotás}\ and\ \citenamefont {Hofer}(2005)}]{Palotas05}%
  \BibitemOpen
  \bibfield  {author} {\bibinfo {author} {\bibfnamefont {K.}~\bibnamefont {Palotás}}\ and\ \bibinfo {author} {\bibfnamefont {W.~A.}\ \bibnamefont {Hofer}},\ }\bibfield  {title} {\bibinfo {title} {Multiple scattering in a vacuum barrier obtained from real-space wavefunctions},\ }\href {https://doi.org/10.1088/0953-8984/17/17/019} {\bibfield  {journal} {\bibinfo  {journal} {J. Phys.: Condens. Matter}\ }\textbf {\bibinfo {volume} {17}},\ \bibinfo {pages} {2705} (\bibinfo {year} {2005})}\BibitemShut {NoStop}%
\bibitem [{\citenamefont {Zhao}\ \emph {et~al.}(2013)\citenamefont {Zhao}, \citenamefont {Levendorf}, \citenamefont {Goncher}, \citenamefont {Schiros}, \citenamefont {Pálová}, \citenamefont {Zabet-Khosousi}, \citenamefont {Rim}, \citenamefont {Gutiérrez}, \citenamefont {Nordlund}, \citenamefont {Jaye}, \citenamefont {Hybertsen}, \citenamefont {Reichman}, \citenamefont {Flynn}, \citenamefont {Park},\ and\ \citenamefont {Pasupathy}}]{Zhao13}%
  \BibitemOpen
  \bibfield  {author} {\bibinfo {author} {\bibfnamefont {L.}~\bibnamefont {Zhao}}, \bibinfo {author} {\bibfnamefont {M.}~\bibnamefont {Levendorf}}, \bibinfo {author} {\bibfnamefont {S.}~\bibnamefont {Goncher}}, \bibinfo {author} {\bibfnamefont {T.}~\bibnamefont {Schiros}}, \bibinfo {author} {\bibfnamefont {L.}~\bibnamefont {Pálová}}, \bibinfo {author} {\bibfnamefont {A.}~\bibnamefont {Zabet-Khosousi}}, \bibinfo {author} {\bibfnamefont {K.~T.}\ \bibnamefont {Rim}}, \bibinfo {author} {\bibfnamefont {C.}~\bibnamefont {Gutiérrez}}, \bibinfo {author} {\bibfnamefont {D.}~\bibnamefont {Nordlund}}, \bibinfo {author} {\bibfnamefont {C.}~\bibnamefont {Jaye}}, \bibinfo {author} {\bibfnamefont {M.}~\bibnamefont {Hybertsen}}, \bibinfo {author} {\bibfnamefont {D.}~\bibnamefont {Reichman}}, \bibinfo {author} {\bibfnamefont {G.~W.}\ \bibnamefont {Flynn}}, \bibinfo {author} {\bibfnamefont {J.}~\bibnamefont {Park}},\ and\ \bibinfo {author} {\bibfnamefont {A.~N.}\ \bibnamefont {Pasupathy}},\ }\bibfield  {title} {\bibinfo
  {title} {Local atomic and electronic structure of boron chemical doping in monolayer graphene},\ }\href {https://doi.org/10.1021/nl401781d} {\bibfield  {journal} {\bibinfo  {journal} {Nano Lett.}\ }\textbf {\bibinfo {volume} {13}},\ \bibinfo {pages} {4659} (\bibinfo {year} {2013})}\BibitemShut {NoStop}%
\bibitem [{\citenamefont {Zheng}\ \emph {et~al.}(2010)\citenamefont {Zheng}, \citenamefont {Hermet},\ and\ \citenamefont {Henrard}}]{Zheng10}%
  \BibitemOpen
  \bibfield  {author} {\bibinfo {author} {\bibfnamefont {B.}~\bibnamefont {Zheng}}, \bibinfo {author} {\bibfnamefont {P.}~\bibnamefont {Hermet}},\ and\ \bibinfo {author} {\bibfnamefont {L.}~\bibnamefont {Henrard}},\ }\bibfield  {title} {\bibinfo {title} {Scanning tunneling microscopy simulations of nitrogen- and boron-doped graphene and single-walled carbon nanotubes},\ }\href {https://doi.org/10.1021/nn1002425} {\bibfield  {journal} {\bibinfo  {journal} {ACS Nano}\ }\textbf {\bibinfo {volume} {4}},\ \bibinfo {pages} {4165} (\bibinfo {year} {2010})}\BibitemShut {NoStop}%
\bibitem [{\citenamefont {Telychko}\ \emph {et~al.}(2014)\citenamefont {Telychko}, \citenamefont {Mutombo}, \citenamefont {Ondráček}, \citenamefont {Hapala}, \citenamefont {Bocquet}, \citenamefont {Kolorenč}, \citenamefont {Vondráček}, \citenamefont {Jelínek},\ and\ \citenamefont {Švec}}]{Telychko14}%
  \BibitemOpen
  \bibfield  {author} {\bibinfo {author} {\bibfnamefont {M.}~\bibnamefont {Telychko}}, \bibinfo {author} {\bibfnamefont {P.}~\bibnamefont {Mutombo}}, \bibinfo {author} {\bibfnamefont {M.}~\bibnamefont {Ondráček}}, \bibinfo {author} {\bibfnamefont {P.}~\bibnamefont {Hapala}}, \bibinfo {author} {\bibfnamefont {F.~C.}\ \bibnamefont {Bocquet}}, \bibinfo {author} {\bibfnamefont {J.}~\bibnamefont {Kolorenč}}, \bibinfo {author} {\bibfnamefont {M.}~\bibnamefont {Vondráček}}, \bibinfo {author} {\bibfnamefont {P.}~\bibnamefont {Jelínek}},\ and\ \bibinfo {author} {\bibfnamefont {M.}~\bibnamefont {Švec}},\ }\bibfield  {title} {\bibinfo {title} {{Achieving high-quality single-atom nitrogen doping of graphene/SiC(0001) by ion implantation and subsequent thermal stabilization}},\ }\href {https://doi.org/10.1021/nn502438k} {\bibfield  {journal} {\bibinfo  {journal} {ACS Nano}\ }\textbf {\bibinfo {volume} {8}},\ \bibinfo {pages} {7318} (\bibinfo {year} {2014})}\BibitemShut {NoStop}%
\bibitem [{\citenamefont {Telychko}\ \emph {et~al.}(2015)\citenamefont {Telychko}, \citenamefont {Mutombo}, \citenamefont {Merino}, \citenamefont {Hapala}, \citenamefont {Ondráček}, \citenamefont {Bocquet}, \citenamefont {Sforzini}, \citenamefont {Stetsovych}, \citenamefont {Vondráček}, \citenamefont {Jelínek},\ and\ \citenamefont {Švec}}]{Telychko15}%
  \BibitemOpen
  \bibfield  {author} {\bibinfo {author} {\bibfnamefont {M.}~\bibnamefont {Telychko}}, \bibinfo {author} {\bibfnamefont {P.}~\bibnamefont {Mutombo}}, \bibinfo {author} {\bibfnamefont {P.}~\bibnamefont {Merino}}, \bibinfo {author} {\bibfnamefont {P.}~\bibnamefont {Hapala}}, \bibinfo {author} {\bibfnamefont {M.}~\bibnamefont {Ondráček}}, \bibinfo {author} {\bibfnamefont {F.~C.}\ \bibnamefont {Bocquet}}, \bibinfo {author} {\bibfnamefont {J.}~\bibnamefont {Sforzini}}, \bibinfo {author} {\bibfnamefont {O.}~\bibnamefont {Stetsovych}}, \bibinfo {author} {\bibfnamefont {M.}~\bibnamefont {Vondráček}}, \bibinfo {author} {\bibfnamefont {P.}~\bibnamefont {Jelínek}},\ and\ \bibinfo {author} {\bibfnamefont {M.}~\bibnamefont {Švec}},\ }\bibfield  {title} {\bibinfo {title} {Electronic and chemical properties of donor, acceptor centers in graphene},\ }\href {https://doi.org/10.1021/acsnano.5b03690} {\bibfield  {journal} {\bibinfo  {journal} {ACS Nano}\ }\textbf {\bibinfo {volume} {9}},\ \bibinfo {pages} {9180} (\bibinfo
  {year} {2015})}\BibitemShut {NoStop}%
\bibitem [{\citenamefont {Tison}\ \emph {et~al.}(2015)\citenamefont {Tison}, \citenamefont {Lagoute}, \citenamefont {Repain}, \citenamefont {Chacon}, \citenamefont {Girard}, \citenamefont {Rousset}, \citenamefont {Joucken}, \citenamefont {Sharma}, \citenamefont {Henrard}, \citenamefont {Amara}, \citenamefont {Ghedjatti},\ and\ \citenamefont {Ducastelle}}]{Tison15}%
  \BibitemOpen
  \bibfield  {author} {\bibinfo {author} {\bibfnamefont {Y.}~\bibnamefont {Tison}}, \bibinfo {author} {\bibfnamefont {J.}~\bibnamefont {Lagoute}}, \bibinfo {author} {\bibfnamefont {V.}~\bibnamefont {Repain}}, \bibinfo {author} {\bibfnamefont {C.}~\bibnamefont {Chacon}}, \bibinfo {author} {\bibfnamefont {Y.}~\bibnamefont {Girard}}, \bibinfo {author} {\bibfnamefont {S.}~\bibnamefont {Rousset}}, \bibinfo {author} {\bibfnamefont {F.}~\bibnamefont {Joucken}}, \bibinfo {author} {\bibfnamefont {D.}~\bibnamefont {Sharma}}, \bibinfo {author} {\bibfnamefont {L.}~\bibnamefont {Henrard}}, \bibinfo {author} {\bibfnamefont {H.}~\bibnamefont {Amara}}, \bibinfo {author} {\bibfnamefont {A.}~\bibnamefont {Ghedjatti}},\ and\ \bibinfo {author} {\bibfnamefont {F.}~\bibnamefont {Ducastelle}},\ }\bibfield  {title} {\bibinfo {title} {Electronic interaction between nitrogen atoms in doped graphene},\ }\href {https://doi.org/10.1021/nn506074u} {\bibfield  {journal} {\bibinfo  {journal} {ACS Nano}\ }\textbf {\bibinfo {volume} {9}},\
  \bibinfo {pages} {670} (\bibinfo {year} {2015})}\BibitemShut {NoStop}%
\bibitem [{\citenamefont {Ferrighi}\ \emph {et~al.}(2015)\citenamefont {Ferrighi}, \citenamefont {Trioni},\ and\ \citenamefont {Di~Valentin}}]{Ferrighi15}%
  \BibitemOpen
  \bibfield  {author} {\bibinfo {author} {\bibfnamefont {L.}~\bibnamefont {Ferrighi}}, \bibinfo {author} {\bibfnamefont {M.~I.}\ \bibnamefont {Trioni}},\ and\ \bibinfo {author} {\bibfnamefont {C.}~\bibnamefont {Di~Valentin}},\ }\bibfield  {title} {\bibinfo {title} {{Boron-doped, nitrogen-doped, and codoped graphene on Cu(111): A DFT + vdW study}},\ }\href {https://doi.org/10.1021/jp512522m} {\bibfield  {journal} {\bibinfo  {journal} {J. Phys. Chem. C}\ }\textbf {\bibinfo {volume} {119}},\ \bibinfo {pages} {6056} (\bibinfo {year} {2015})}\BibitemShut {NoStop}%
\bibitem [{\citenamefont {Li}\ \emph {et~al.}(2018)\citenamefont {Li}, \citenamefont {Lin}, \citenamefont {Lv}, \citenamefont {Terrones}, \citenamefont {Chi}, \citenamefont {Hofer},\ and\ \citenamefont {Pan}}]{Li18}%
  \BibitemOpen
  \bibfield  {author} {\bibinfo {author} {\bibfnamefont {Q.}~\bibnamefont {Li}}, \bibinfo {author} {\bibfnamefont {H.}~\bibnamefont {Lin}}, \bibinfo {author} {\bibfnamefont {R.}~\bibnamefont {Lv}}, \bibinfo {author} {\bibfnamefont {M.}~\bibnamefont {Terrones}}, \bibinfo {author} {\bibfnamefont {L.}~\bibnamefont {Chi}}, \bibinfo {author} {\bibfnamefont {W.~A.}\ \bibnamefont {Hofer}},\ and\ \bibinfo {author} {\bibfnamefont {M.}~\bibnamefont {Pan}},\ }\bibfield  {title} {\bibinfo {title} {Locally induced spin states on graphene by chemical attachment of boron atoms},\ }\href {https://doi.org/10.1021/acs.nanolett.8b01798} {\bibfield  {journal} {\bibinfo  {journal} {Nano Lett.}\ }\textbf {\bibinfo {volume} {18}},\ \bibinfo {pages} {5482} (\bibinfo {year} {2018})}\BibitemShut {NoStop}%
\bibitem [{\citenamefont {Joucken}\ \emph {et~al.}(2019)\citenamefont {Joucken}, \citenamefont {Henrard},\ and\ \citenamefont {Lagoute}}]{Joucken19}%
  \BibitemOpen
  \bibfield  {author} {\bibinfo {author} {\bibfnamefont {F.}~\bibnamefont {Joucken}}, \bibinfo {author} {\bibfnamefont {L.}~\bibnamefont {Henrard}},\ and\ \bibinfo {author} {\bibfnamefont {J.}~\bibnamefont {Lagoute}},\ }\bibfield  {title} {\bibinfo {title} {Electronic properties of chemically doped graphene},\ }\href {https://doi.org/10.1103/PhysRevMaterials.3.110301} {\bibfield  {journal} {\bibinfo  {journal} {Phys. Rev. Mater.}\ }\textbf {\bibinfo {volume} {3}},\ \bibinfo {pages} {110301} (\bibinfo {year} {2019})}\BibitemShut {NoStop}%
\bibitem [{\citenamefont {Neilson}\ \emph {et~al.}(2019)\citenamefont {Neilson}, \citenamefont {Chinkezian}, \citenamefont {Phirke}, \citenamefont {Osei-Twumasi}, \citenamefont {Li}, \citenamefont {Chichiri}, \citenamefont {Cho}, \citenamefont {Palotás}, \citenamefont {Gan}, \citenamefont {Garrett}, \citenamefont {Lau},\ and\ \citenamefont {Gao}}]{Neilson19}%
  \BibitemOpen
  \bibfield  {author} {\bibinfo {author} {\bibfnamefont {J.}~\bibnamefont {Neilson}}, \bibinfo {author} {\bibfnamefont {H.}~\bibnamefont {Chinkezian}}, \bibinfo {author} {\bibfnamefont {H.}~\bibnamefont {Phirke}}, \bibinfo {author} {\bibfnamefont {A.}~\bibnamefont {Osei-Twumasi}}, \bibinfo {author} {\bibfnamefont {Y.}~\bibnamefont {Li}}, \bibinfo {author} {\bibfnamefont {C.}~\bibnamefont {Chichiri}}, \bibinfo {author} {\bibfnamefont {J.}~\bibnamefont {Cho}}, \bibinfo {author} {\bibfnamefont {K.}~\bibnamefont {Palotás}}, \bibinfo {author} {\bibfnamefont {L.}~\bibnamefont {Gan}}, \bibinfo {author} {\bibfnamefont {S.~J.}\ \bibnamefont {Garrett}}, \bibinfo {author} {\bibfnamefont {K.~C.}\ \bibnamefont {Lau}},\ and\ \bibinfo {author} {\bibfnamefont {L.}~\bibnamefont {Gao}},\ }\bibfield  {title} {\bibinfo {title} {Nitrogen-doped graphene on copper: Edge-guided doping process and doping-induced variation of local work function},\ }\href {https://doi.org/10.1021/acs.jpcc.8b11261} {\bibfield  {journal} {\bibinfo
  {journal} {J. Phys. Chem. C}\ }\textbf {\bibinfo {volume} {123}},\ \bibinfo {pages} {8802} (\bibinfo {year} {2019})}\BibitemShut {NoStop}%
\bibitem [{\citenamefont {Yang}\ \emph {et~al.}(2023)\citenamefont {Yang}, \citenamefont {Abilio}, \citenamefont {Bernal~Romero}, \citenamefont {Rodriguez}, \citenamefont {Escobar~Godoy}, \citenamefont {Little}, \citenamefont {Mckee}, \citenamefont {Carbajal}, \citenamefont {Li}, \citenamefont {Chen}, \citenamefont {Gao}, \citenamefont {Palotás},\ and\ \citenamefont {Gao}}]{Yang23}%
  \BibitemOpen
  \bibfield  {author} {\bibinfo {author} {\bibfnamefont {H.}~\bibnamefont {Yang}}, \bibinfo {author} {\bibfnamefont {I.}~\bibnamefont {Abilio}}, \bibinfo {author} {\bibfnamefont {J.}~\bibnamefont {Bernal~Romero}}, \bibinfo {author} {\bibfnamefont {C.}~\bibnamefont {Rodriguez}}, \bibinfo {author} {\bibfnamefont {M.}~\bibnamefont {Escobar~Godoy}}, \bibinfo {author} {\bibfnamefont {M.}~\bibnamefont {Little}}, \bibinfo {author} {\bibfnamefont {P.}~\bibnamefont {Mckee}}, \bibinfo {author} {\bibfnamefont {V.}~\bibnamefont {Carbajal}}, \bibinfo {author} {\bibfnamefont {J.}~\bibnamefont {Li}}, \bibinfo {author} {\bibfnamefont {X.}~\bibnamefont {Chen}}, \bibinfo {author} {\bibfnamefont {H.-J.}\ \bibnamefont {Gao}}, \bibinfo {author} {\bibfnamefont {K.}~\bibnamefont {Palotás}},\ and\ \bibinfo {author} {\bibfnamefont {L.}~\bibnamefont {Gao}},\ }\bibfield  {title} {\bibinfo {title} {{Atomic-scale identification of nitrogen dopants in graphene on Ir(111) and Ru(0001)}},\ }\href {https://doi.org/10.1088/1361-648X/ace229}
  {\bibfield  {journal} {\bibinfo  {journal} {J. Phys.: Condens. Matter}\ }\textbf {\bibinfo {volume} {35}},\ \bibinfo {pages} {405003} (\bibinfo {year} {2023})}\BibitemShut {NoStop}%
\bibitem [{\citenamefont {Mándi}\ \emph {et~al.}(2013)\citenamefont {Mándi}, \citenamefont {Nagy},\ and\ \citenamefont {Palotás}}]{Mandi13}%
  \BibitemOpen
  \bibfield  {author} {\bibinfo {author} {\bibfnamefont {G.}~\bibnamefont {Mándi}}, \bibinfo {author} {\bibfnamefont {N.}~\bibnamefont {Nagy}},\ and\ \bibinfo {author} {\bibfnamefont {K.}~\bibnamefont {Palotás}},\ }\bibfield  {title} {\bibinfo {title} {{Arbitrary tip orientation in STM simulations: 3D WKB theory and application to W(110)}},\ }\href {https://doi.org/10.1088/0953-8984/25/44/445009} {\bibfield  {journal} {\bibinfo  {journal} {J. Phys.: Condens. Matter}\ }\textbf {\bibinfo {volume} {25}},\ \bibinfo {pages} {445009} (\bibinfo {year} {2013})}\BibitemShut {NoStop}%
\bibitem [{\citenamefont {Mándi}\ \emph {et~al.}(2014)\citenamefont {Mándi}, \citenamefont {Teobaldi},\ and\ \citenamefont {Palotás}}]{Mandi14}%
  \BibitemOpen
  \bibfield  {author} {\bibinfo {author} {\bibfnamefont {G.}~\bibnamefont {Mándi}}, \bibinfo {author} {\bibfnamefont {G.}~\bibnamefont {Teobaldi}},\ and\ \bibinfo {author} {\bibfnamefont {K.}~\bibnamefont {Palotás}},\ }\bibfield  {title} {\bibinfo {title} {{Contrast stability and ‘stripe’ formation in scanning tunnelling microscopy imaging of highly oriented pyrolytic graphite: the role of STM-tip orientations}},\ }\href {https://doi.org/10.1088/0953-8984/26/48/485007} {\bibfield  {journal} {\bibinfo  {journal} {J. Phys.: Condens. Matter}\ }\textbf {\bibinfo {volume} {26}},\ \bibinfo {pages} {485007} (\bibinfo {year} {2014})}\BibitemShut {NoStop}%
\bibitem [{\citenamefont {Mándi}\ \emph {et~al.}(2015)\citenamefont {Mándi}, \citenamefont {Teobaldi},\ and\ \citenamefont {Palotás}}]{Mandi15progsurfsci}%
  \BibitemOpen
  \bibfield  {author} {\bibinfo {author} {\bibfnamefont {G.}~\bibnamefont {Mándi}}, \bibinfo {author} {\bibfnamefont {G.}~\bibnamefont {Teobaldi}},\ and\ \bibinfo {author} {\bibfnamefont {K.}~\bibnamefont {Palotás}},\ }\bibfield  {title} {\bibinfo {title} {What is the orientation of the tip in a scanning tunneling microscope?},\ }\href {https://doi.org/https://doi.org/10.1016/j.progsurf.2015.02.001} {\bibfield  {journal} {\bibinfo  {journal} {Prog. Surf. Sci.}\ }\textbf {\bibinfo {volume} {90}},\ \bibinfo {pages} {223} (\bibinfo {year} {2015})}\BibitemShut {NoStop}%
\bibitem [{\citenamefont {Choudhary}\ \emph {et~al.}(2021)\citenamefont {Choudhary}, \citenamefont {Garrity}, \citenamefont {Camp}, \citenamefont {Kalinin}, \citenamefont {Vasudevan}, \citenamefont {Ziatdinov},\ and\ \citenamefont {Tavazza}}]{Choudhary21}%
  \BibitemOpen
  \bibfield  {author} {\bibinfo {author} {\bibfnamefont {K.}~\bibnamefont {Choudhary}}, \bibinfo {author} {\bibfnamefont {K.~F.}\ \bibnamefont {Garrity}}, \bibinfo {author} {\bibfnamefont {C.}~\bibnamefont {Camp}}, \bibinfo {author} {\bibfnamefont {S.~V.}\ \bibnamefont {Kalinin}}, \bibinfo {author} {\bibfnamefont {R.}~\bibnamefont {Vasudevan}}, \bibinfo {author} {\bibfnamefont {M.}~\bibnamefont {Ziatdinov}},\ and\ \bibinfo {author} {\bibfnamefont {F.}~\bibnamefont {Tavazza}},\ }\bibfield  {title} {\bibinfo {title} {Computational scanning tunneling microscope image database},\ }\href {https://doi.org/10.1038/s41597-021-00824-y} {\bibfield  {journal} {\bibinfo  {journal} {Sci. Data}\ }\textbf {\bibinfo {volume} {8}},\ \bibinfo {pages} {57} (\bibinfo {year} {2021})}\BibitemShut {NoStop}%
\bibitem [{\citenamefont {Guerrero-Rivera}\ \emph {et~al.}(2024)\citenamefont {Guerrero-Rivera}, \citenamefont {Godínez-Garcia}, \citenamefont {Hayashi}, \citenamefont {Wang},\ and\ \citenamefont {Ortiz-Medina}}]{Guerrero-Rivera24}%
  \BibitemOpen
  \bibfield  {author} {\bibinfo {author} {\bibfnamefont {R.}~\bibnamefont {Guerrero-Rivera}}, \bibinfo {author} {\bibfnamefont {F.~J.}\ \bibnamefont {Godínez-Garcia}}, \bibinfo {author} {\bibfnamefont {T.}~\bibnamefont {Hayashi}}, \bibinfo {author} {\bibfnamefont {Z.}~\bibnamefont {Wang}},\ and\ \bibinfo {author} {\bibfnamefont {J.}~\bibnamefont {Ortiz-Medina}},\ }\bibfield  {title} {\bibinfo {title} {{Machine-learning driven STM images prediction of doped/defective graphene: Towards optimized tools for 2D nanomaterials characterization}},\ }\href {https://doi.org/https://doi.org/10.1016/j.commatsci.2024.113076} {\bibfield  {journal} {\bibinfo  {journal} {Comput. Mater. Sci.}\ }\textbf {\bibinfo {volume} {242}},\ \bibinfo {pages} {113076} (\bibinfo {year} {2024})}\BibitemShut {NoStop}%
\bibitem [{\citenamefont {Kresse}\ and\ \citenamefont {Furthm\"uller}(1996)}]{vasp}%
  \BibitemOpen
  \bibfield  {author} {\bibinfo {author} {\bibfnamefont {G.}~\bibnamefont {Kresse}}\ and\ \bibinfo {author} {\bibfnamefont {J.}~\bibnamefont {Furthm\"uller}},\ }\bibfield  {title} {\bibinfo {title} {Efficient iterative schemes for ab initio total-energy calculations using a plane-wave basis set},\ }\href {https://doi.org/10.1103/PhysRevB.54.11169} {\bibfield  {journal} {\bibinfo  {journal} {Phys. Rev. B}\ }\textbf {\bibinfo {volume} {54}},\ \bibinfo {pages} {11169} (\bibinfo {year} {1996})}\BibitemShut {NoStop}%
\bibitem [{\citenamefont {Perdew}\ and\ \citenamefont {Wang}(1992)}]{PW91}%
  \BibitemOpen
  \bibfield  {author} {\bibinfo {author} {\bibfnamefont {J.~P.}\ \bibnamefont {Perdew}}\ and\ \bibinfo {author} {\bibfnamefont {Y.}~\bibnamefont {Wang}},\ }\bibfield  {title} {\bibinfo {title} {Accurate and simple analytic representation of the electron-gas correlation energy},\ }\href {https://doi.org/10.1103/PhysRevB.45.13244} {\bibfield  {journal} {\bibinfo  {journal} {Phys. Rev. B}\ }\textbf {\bibinfo {volume} {45}},\ \bibinfo {pages} {13244} (\bibinfo {year} {1992})}\BibitemShut {NoStop}%
\bibitem [{\citenamefont {Kresse}\ and\ \citenamefont {Joubert}(1999)}]{paw}%
  \BibitemOpen
  \bibfield  {author} {\bibinfo {author} {\bibfnamefont {G.}~\bibnamefont {Kresse}}\ and\ \bibinfo {author} {\bibfnamefont {D.}~\bibnamefont {Joubert}},\ }\bibfield  {title} {\bibinfo {title} {From ultrasoft pseudopotentials to the projector augmented-wave method},\ }\href {https://doi.org/10.1103/PhysRevB.59.1758} {\bibfield  {journal} {\bibinfo  {journal} {Phys. Rev. B}\ }\textbf {\bibinfo {volume} {59}},\ \bibinfo {pages} {1758} (\bibinfo {year} {1999})}\BibitemShut {NoStop}%
\bibitem [{\citenamefont {Monkhorst}\ and\ \citenamefont {Pack}(1976)}]{Monkhorst76}%
  \BibitemOpen
  \bibfield  {author} {\bibinfo {author} {\bibfnamefont {H.~J.}\ \bibnamefont {Monkhorst}}\ and\ \bibinfo {author} {\bibfnamefont {J.~D.}\ \bibnamefont {Pack}},\ }\bibfield  {title} {\bibinfo {title} {{Special points for Brillouin-zone integrations}},\ }\href {https://doi.org/10.1103/PhysRevB.13.5188} {\bibfield  {journal} {\bibinfo  {journal} {Phys. Rev. B}\ }\textbf {\bibinfo {volume} {13}},\ \bibinfo {pages} {5188} (\bibinfo {year} {1976})}\BibitemShut {NoStop}%
\bibitem [{\citenamefont {Teobaldi}\ \emph {et~al.}(2012)\citenamefont {Teobaldi}, \citenamefont {Inami}, \citenamefont {Kanasaki}, \citenamefont {Tanimura},\ and\ \citenamefont {Shluger}}]{Teobaldi12}%
  \BibitemOpen
  \bibfield  {author} {\bibinfo {author} {\bibfnamefont {G.}~\bibnamefont {Teobaldi}}, \bibinfo {author} {\bibfnamefont {E.}~\bibnamefont {Inami}}, \bibinfo {author} {\bibfnamefont {J.}~\bibnamefont {Kanasaki}}, \bibinfo {author} {\bibfnamefont {K.}~\bibnamefont {Tanimura}},\ and\ \bibinfo {author} {\bibfnamefont {A.~L.}\ \bibnamefont {Shluger}},\ }\bibfield  {title} {\bibinfo {title} {Role of applied bias and tip electronic structure in the scanning tunneling microscopy imaging of highly oriented pyrolytic graphite},\ }\href {https://doi.org/10.1103/PhysRevB.85.085433} {\bibfield  {journal} {\bibinfo  {journal} {Phys. Rev. B}\ }\textbf {\bibinfo {volume} {85}},\ \bibinfo {pages} {085433} (\bibinfo {year} {2012})}\BibitemShut {NoStop}%
\end{thebibliography}

%

\end{document}